\documentstyle[12pt]{article}
\def\myrm{\rm}

\def\setfonts{%
\font\frbig=eufm10 scaled\magstep1
\font\frscr=eufm10
\font\frscrscr=eufm8
\newfam\frfam
\textfont\frfam=\frbig
\scriptfont\frfam=\frscr
\scriptscriptfont\frfam=\frscrscr
\def\fr{\fam\frfam}
\font\openbig=msbm10 scaled\magstep1
\font\openscr=msbm10
\font\openscrscr=msbm8
\newfam\openfam
\textfont\openfam=\openbig
\scriptfont\openfam=\openscr
\scriptscriptfont\openfam=\openscrscr
\def\open{\fam\openfam}
}

\setfonts

\font\tenrm  = cmr10
\font\tenmi  = cmmi10
  \skewchar\tenmi ='177
\font\tensy  = cmsy10
  \skewchar\tensy ='60
\font\tenex  = cmex10

\font\tensf  = cmss10
\font\tenly  = lasy10

\font\egtsf=cmss8

\font\sevrm  = cmr7
\font\sevmi  = cmmi7
  \skewchar\sevmi ='177
\font\sevsy  = cmsy7
  \skewchar\sevsy ='60

\font\fivrm  = cmr5
\font\fivmi  = cmmi5
  \skewchar\fivmi ='177
\font\fivsy  = cmsy5
  \skewchar\fivsy ='60

\makeatletter

\def\tableofcontents{\subsection*{\mbox{}
} \@starttoc{toc}}
\newdimen\normalarrayskip
\newdimen\minarrayskip
\normalarrayskip\baselineskip
\minarrayskip\jot
\newif\ifold \oldtrue \def\new{\oldfalse}
\def\arraymode{\ifold\relax\else\displaystyle\fi}

\def\@arrayskip{\ifold\baselineskip\z@\lineskip\z@
  \else
  \baselineskip\minarrayskip\lineskip2\minarrayskip\fi}
\def\@arrayclassz{\ifcase \@lastchclass \@acolampacol \or
\@ampacol \or \or \or \@addamp \or
 \@acolampacol \or \@firstampfalse \@acol \fi
\edef\@preamble{\@preamble
 \ifcase \@chnum
  \hfil$\relax\arraymode\@sharp$\hfil
  \or $\relax\arraymode\@sharp$\hfil
  \or \hfil$\relax\arraymode\@sharp$\fi}}
\def\@array[#1]#2{\setbox\@arstrutbox=\hbox{\vrule
  height\arraystretch \ht\strutbox
  depth\arraystretch \dp\strutbox
  width\z@}\@mkpream{#2}\edef\@preamble{\halign \noexpand\@halignto
\bgroup \tabskip\z@ \@arstrut \@preamble \tabskip\z@ \cr}%
\let\@startpbox\@@startpbox \let\@endpbox\@@endpbox
 \if #1t\vtop \else \if#1b\vbox \else \vcenter \fi\fi
 \bgroup \let\par\relax
 \let\@sharp##\let\protect\relax
 \@arrayskip\@preamble}
\def\l@section#1#2{\addpenalty{\@secpenalty} \addvspace{.4em plus 1pt}
\@tempdima 1.5em \begingroup
 \parindent \z@ \rightskip \@pnumwidth
 \parfillskip -\@pnumwidth
 \small\sf \leavevmode \advance\leftskip\@tempdima
 \hskip -\leftskip #1\nobreak\hfil
\nobreak\hbox to\@pnumwidth{\hss #2}\par
 \endgroup}
\def\tenpt{%
\textfont\z@\tenrm
  \scriptfont\z@\sevrm \scriptscriptfont\z@\fivrm
\textfont\@ne\tenmi \scriptfont\@ne\sevmi \scriptscriptfont\@ne\fivmi
\textfont\tw@\tensy \scriptfont\tw@\sevsy \scriptscriptfont\tw@\fivsy
\textfont\thr@@\tenex \scriptfont\thr@@\tenex \scriptscriptfont\thr@@\tenex
\def\unboldmath{\everymath{}\everydisplay{}\@nomath\unboldmath
  \textfont\@ne\tenmi
  \textfont\tw@\tensy \textfont\lyfam\tenly
  \@boldfalse}\@boldfalse
\def\boldmath{\@ifundefined{tenmib}{\global\font\tenmib\@mbi
   \global\font\tensyb\@mbsy
   \global\font\tenlyb\@lasyb\relax\@addfontinfo\@xpt
   {\def\boldmath{\everymath{\mit}\everydisplay{\mit}\@prtct\@nomathbold
  \textfont\@ne\tenmib \textfont\tw@\tensyb
  \textfont\lyfam\tenlyb \@prtct\@boldtrue}}}{}\@xpt\boldmath}%
\def\psf{\fam\sffam\tensf}\textfont\sffam\tensf
  \scriptfont\sffam\egtsf \scriptscriptfont\sffam\egtsf
}
\makeatother

\def\theequation{\thesection.\arabic{equation}}
\let\ssection=\section
\def\section{\setcounter{equation}{0}\ssection}

\def\lvm{\leavevmode\hbox to\parindent{\hfill}}
\def\req#1{(\ref{#1})}

\def\BE{\begin{equation}}
\def\EE{\end{equation} }
\def\BA{\begin{array}}
\def\EA{\end{array}}
\def\BB{\begin{equation}\new\begin{array}{rcl}}
\def\L{\left}
\def\R{\right}

\def\bar{\overline}
\def\frac#1#2{{\textstyle{{#1}\over{#2}}}}
\def\Kr#1{\delta_{{#1},0}}
\def\ket#1{\bigl|{#1}\bigr\rangle}

\def\d{\partial}

\def\half{{\textstyle{1\over2}}}
\def\third{{\textstyle{1\over3}}}
\def\fourth{{\textstyle{1\over4}}}
\def\sixth{{\textstyle{1\over6}}}
\def\eighth{{\textstyle{1\over8}}}

\def\a{\alpha}

\def\cA{{\cal A}}
\def\cB{{\cal B}}
\def\cC{{\cal C}}
\def\cG{{\cal G}}
\def\cH{{\cal H}}
\def\cL{{\cal L}}
\def\cO{{\cal O}}
\def\cQ{{\cal Q}}
\def\cT{{\cal T}}
\def\cU{{\cal U}}

\def\oZ{{\open Z}}

\def\tLS{{\widetilde{L}}^{{\myrm{S}}}}

\def\frt{{\fr t}}
\def\frm{{\fr m}}
\def\frl{{\fr l}}

\def\ctop{{\sf c}}
\def\htop{{\sf h}}

\def\sL{{\sf L}}
\def\sQ{{\sf Q}}

\def\bc{background charge}
\def\cc{central charge}
\def\tcc{topological central charge}
\def\emt{energy-momentum tensor}

\begin{document}
\hfuzz=1.8pt
\raggedbottom
\begin{flushright}
{\tt hep-th@xxx}/9311180
\end{flushright}
\mbox{}
\vskip12ex
\begin{center}\Large\sc The MFF Singular Vectors in Topological\\
Conformal Theories
\end{center}

\medskip

\centerline{\large A.~M.~Semikhatov}
\vskip0.5ex
\begin{center}
{\small\sl I.E.~Tamm Theory Division, P.~N.~Lebedev Physics Institute\\
Russian Academy of Sciences, 53 Leninski prosp., Moscow 117924\\
Russia}
\end{center}
\vskip4ex
\centerline{\sc abstract}
\vskip1.2ex
\centerline{\parbox{.9\hsize}{\addtolength\baselineskip{-2ex}
\noindent\small
It is argued that singular vectors of the topological conformal (twisted
$N\!=\!2$) algebra are identical with singular vectors of the $sl(2)$
Ka\v{c}--Moody algebra. An arbitrary matter theory can be dressed by
additional fields to make up a representation of either the $sl(2)$ current
algebra or the topological conformal algebra. The relation between the two
constructions is equivalent to the Kazama--Suzuki realisation of a
topological conformal theory as $sl(2)\oplus u(1)/u(1)$. The
Malikov--Feigin--Fuchs (MFF) formula for the $sl(2)$ singular vectors
translates into a general expression for the topological singular vectors.
The MFF/topological singular states are observed to vanish in
Witten's free-field construction of the (twisted) $N\!=\!2$ algebra,
derived from the Landau--Ginzburg formalism.}}

\thispagestyle{empty}

\bigskip

\tableofcontents

\newpage
\setcounter{page}{1}
\section{Introduction}\lvm Associated with singular (`null') states in the
representations of symmetry algebras encountered in conformal theories, are
the irreducible models whose correlators satisfy decoupling equations -- i.e.
the equations ensuring the decoupling of a given null vector. The decoupling
equations do in fact define a class of conformal models
\cite{[BPZ],[DF],[FQS]}. Besides, singular vectors and their decoupling
equations have been used to relate conformal field-theoretic constructions to
other structures in physics and mathematics
\cite{[dFM],[GoSi],[P],Solving,[GS3]}. Singular vectors of the Virasoro
algebra (corresponding to the {\sl minimal\/} models) and its supersymmetric
extensions have been studied in some detail
\cite{[FF],[BSa],[BdFIZ],[BS],[Kent],[GP],[MRc],[BFK],[Kir]}, as were
singular vectors of the $sl(2)$ Ka\v{c}--Moody algebra \cite{[KK],[MFF]} and
their relation with the Virasoro ones \cite{[BdFIZ]}--\cite{[GP1]}.

In this paper we consider singular vectors of the $N\!=\!2$ supersymmetry
algebra in its {\it topological\/} (twisted) version \cite{[Ey],[W-top]}.
The `topological' singular vectors are interesting for a number of (not
independent) reasons. First, as we will see below, they can be identified
with singular vectors of $sl(2)$ Ka\v{c}--Moody algebra (the latter being
given explicitly by the Malikov--Feigin--Fuchs (MFF) construction
\cite{[MFF]}). At the same time they have been related \cite{Solving,[GS3]}
to Virasoro constraints of the type of those encountered in matrix models.
Both these features have to do with the fact that topological conformal
theories can be reduced to ordinary minimal matter in such a way that the
topological singular vectors map into the `minimal' ones. Conversely, it is
possible to represent a topological conformal model as a result of dressing
the minimal matter with (free) gravity multiplet (`Liouville' and ghosts). A
very interesting generalisation of this and similar constructions consists in
the idea \cite{[BV]} of universal string theory \cite{[BV],[Fof],[IK]}.
Moreover, a similar correspondence might exist between ordinary matter
theories and $sl(2)$ WZW models (based on the fact that
the $sl(2)$ algebra can be built by dressing
an ordinary matter theory). The hidden connection, so far observed at the
level of singular vectors, between the topological and the $sl(2)$ algebras,
may be interesting, in particular, in the context of universal string theory.

\medskip

The topological conformal algebra with \tcc\ $\ctop$,\BE\new\BA{lclclcl}
\L[\cL_m,\cL_n\R]&=&(m-n)\cL_{m+n}\,,&\qquad&[\cH_m,\cH_n]&=
&{\ctop\over3}m\Kr{m+n}\,,\\
\L[\cL_m,\cG_n\R]&=&(m-n)\cG_{m+n}\,,&\qquad&[\cH_m,\cG_n]&=&\cG_{m+n}\,,\\
\L[\cL_m,\cQ_n\R]&=&-n\cQ_{m+n}\,,&\qquad&[\cH_m,\cQ_n]&=&-\cQ_{m+n}\,,\\
\L[\cL_m,\cH_n\R]&=&\multicolumn{5}{l}{-n\cH_{m+n}+{\ctop\over6}(m^2+m)
\Kr{m+n}\,,}\\
\L\{\cG_m,\cQ_n\R\}&=&\multicolumn{5}{l}{2\cL_{m+n}-2n\cH_{m+n}+
{\ctop\over3}(m^2+m)\Kr{m+n}\,,}\EA\qquad m,~n\in\oZ\,.\label{topalgebra}\EE
admits, for $\ctop\neq3$, a construction \cite{[GS2],[GS3],[BLNW],[MV],[GR]}
in terms of ordinary (non-supersymmetric) matter with \cc\BE
d={(\ctop+1)(\ctop+6)\over\ctop-3}\,,\label{d(c)}\EE an auxiliary scalar
field, and a couple of $bc$ ghosts. To be more precise, two such
constructions exist \cite{[GS2],[GS3]}, one of which reproduces in the
(matter) + (scalar $\phi$) sector the DDK recipe \cite{[Da],[DK]} for
dressing matter with the Liouville, which will allow us to call the $\phi$
scalar the Liouville. The second way to build up the topological algebra out
of \cc-$d$ matter, ghosts, and the Liouville involves ghosts of dimension
(`spin') 1 (rather than 2 as in the DDK-related version). This will be
referred to as the `mirror' gravity coupled to matter, and it this version
that is interesting in application to singular states. The two versions
follow as different twistings \cite{[D-pr],[GS3]} of a free-field realisation
of the proper $N\!=\!2$ algebra.

The topological singular states, which we are going to study, are built upon
chiral primary states \cite{[LVW]} $\ket\Psi\equiv\ket\htop$ of the algebra
\req{topalgebra}:
\BE\new\BA{rcl}\cQ_0\ket\htop&=&0\,,\qquad\cG_0\ket\htop~{}={}~0,\qquad
\cH_0\ket\htop~{}={}~\htop\ket\htop\,,\\
\cL_{\geq0}\ket\htop&=&\cH_{\geq1}\ket\htop~{}={}~
\cG_{\geq1}\ket\htop~{}={}~\cQ_{\geq1}\ket\htop~{}={}~0\,,
\EA\label{cpsconditions}\EE (the topological $U(1)$ charge $\htop$ being the
only parameter that distinguishes between chiral primary states). Singular
vectors are those descendants of $\ket\htop$ that satisfy the highest-weight
conditions \req{cpsconditions} except for the chirality ($\cG_0$) and
zero-dimension ($\cL_0$). In the following, we will show that they coincide
with the $sl(2)$ Ka\v{c}--Moody singular states. This points to a closer
correspondence between the topological and WZW theories, as is also suggested
by a number of recent results \cite{[ASy],[Hy],[Sv]}, and might also be
interpreted in terms of the appearance of the $sl(2)$ symmetry in quantum
gravity \cite{[Po]}.

\medskip

In the next section we recall a free-field construction of the algebra
\req{topalgebra} in terms of matter, `Liouville', and ghosts.  Then, in
section~3, we explicitly construct several lowest topological singular
vectors. Section~4 contains the necessary preparations for the subsequent
evaluation of these singular vectors in terms of the $sl(2)$ algebra. In
section~5 we point out a construction, analogous to the one producing
topological algebra via dressing ordinary matter, that allows us to build up
the $sl(2)$ currents out of ordinary matter and a couple of scalars. This
construction is basically a variation on the theme of hamiltonian reduction
and Wakimoto bosonisation, but we would like to stress that the matter enters
only through its Virasoro generators and need not be specified any further.
In section~6 we recall briefly what the singular vectors of $sl(2)$ look like
(the Malikov--Feigin--Fuchs formulae \cite{[MFF]}). Finally, in section~7 we
state the main result, namely that the topological vectors from section~3,
when evaluated according to the recipe given in section~4, become exactly the
MFF states of section~6. In fact, the topological singular states can also be
used to `resolve' the MFF ones in the classical limit $|k|=\infty$, as
discussed in section~8, where we also make comparison with a different kind
of a free-field construction for the topological algebra, proposed recently
by Witten \cite{[W-LG]} on the basis of the Landau--Ginzburg formalism. In
section~9 we use this free-field construction to evaluate the MFF/topological
singular states in terms of the Landau--Ginzburg theory. Section~10 contains
several concluding remarks.

\pagebreak[3]

\section{Topological algebra: free-field constructions}\lvm Both the `mirror'
and the `ordinary' versions of the construction presenting the topological
algebra as matter dressed with the gravity multiplet, can be obtained by
performing the two possible twistings of a similar realisation for the
proper (untwisted) $N\!=\!2$ algebra\BE\new\BA{lclclcl}
\L[\cL_m,\cL_n\R]&=&(m-n)\cL_{m+n}+{\ctop\over12}(m^3-m)\Kr{m+n}
\,,&\qquad&[\cH_m,\cH_n]&=&{\ctop\over3}m\Kr{m+n}\,,\\ \L[\cL_m,\cG_r^\pm
\R]&=&\L({m\over2}-r\R)\cG_{m+r}^\pm
\,,&\qquad&[\cH_m,\cG_r^\pm]&=&\pm\cG_{m+r}^\pm\,,\\
\L[\cL_m,\cH_n\R]&=&{}-n\cH_{m+n}\,,&{}&\multicolumn{3}{r}{
m,\,n\in\oZ\,,\qquad r,s\in\oZ+\half}\\
\L\{\cG_r^-,\cG_s^+\R\}&=&\multicolumn{5}{l}{2\cL_{r+s}-(r-s)\cH_{r+s}+
{\ctop\over3}(r^2-\frac{1}{4})\Kr{r+s}\,,}\EA\label{N2algebra}\EE For this
algebra, a representation in terms of a central-charge-$d$ matter,
spin-$3\over3$ $bc$ ghosts, and an extra scalar current $I=\d\phi$, reads
\cite{[GS3]}\BB \cT(z)&=&{}T(z)-\half I(z)^2-\fourth(Q_{{\myrm{L}}}+Q)\d
I(z)+t(z)\,,\\
\cH(z)&=&{}i(z)-\half(Q_{{\myrm{L}}}-Q)I(z)\,,\qquad\cG^-(z)~=~b(z)\,,\\
\cG^+(z)&=&{}2c(z)T(z)-2b(z)\d c(z)\!\cdot\! c(z)+(Q_{{\myrm{L}}}-Q)\d
c(z)\!\cdot\!I(z) -Qc(z)\d I(z)+{\ctop\over3}\d^2c(z)
\EA\label{construction}\EE where $T$ and $t$ are the matter and the ghost
\emt s respectively, while $i$ is the ghost current $i=-bc$:\BE
i(z)i(w)={1\over(z-w)^2},\qquad I(z)I(w)={-1\over(z-w)^2}\label{signatures}
\EE and (for $\ctop\!<\!3$ for definiteness)\BE
Q={\ctop+3\over\sqrt{3(3-\ctop)}}\,,\qquad
Q_{{\myrm{L}}}={9-\ctop\over\sqrt{3(3-\ctop)}}\,,\EE The crucial operator
product\BE\cG^-(z)\cG^+(w)={2\ctop\over3}{1\over(z-w)^3}+
{-\cH(w)-\cH(z)\over(z-w)^2}+{2\cT\over z-w}\,,\EE as well as the other
relations of the algebra \req{N2algebra} can now be verified provided $d$ is
related to the $N\!=\!2$ \cc\ $\ctop$ via eq.~\req{d(c)}.

\smallskip

There are just {\it two\/} possibilities to twist the algebra
\req{N2algebra}, with either $\cG^+$ or $\cG^-$ acquiring spin 2 after
twisting. The first twisting is accomplished by setting\BE\new\BA{rclcrcl}
\cL^{(1)}_m&=&\multicolumn{5}{l}{\cL_m+\half(m+1)\cH_m\,,}\\
c^{(1)}_m&=&c_{m+\half}\,,&\qquad &b^{(1)}_m&=&b_{m-\half}\,,\\
\cH^{(1)}_m&=&\cH_m\,,&{}&{}&{}&{}\\ \cG^{(1)}_m&=&\cG_{m-\half}^+\,,&\qquad
&\cQ_n^{(1)}&=&\cG^-_{n+\half}\,,\EA\label{spin1}\EE which gives a `spin-1'
(for the spin of the ghosts $b^{(1)}c^{(1)}$) construction of the topological
algebra. We thus find \BB\cT^{(1)}(z)&=&\cT(z)-\half\d\cH(z)\\
{}&=&{}T(z)-\half I(z)^2+t(z)-\half Q\d I(z)-\half\d
i(z)\,.\EA\label{construction1}\EE which renders the \bc\ of the current $I$
the same as the matter \bc\ $Q$. As the topological $U(1)$ current $\cH$
contains the ghost current, the ghosts' dimensions are also shifted, by the
$\d i$-term, to 1 and 0 for $b$ and $c$ respectively, producing the ghost
\emt\BE t^{(1)}=t-\half\d i=-b\d c\,.\EE The resulting $bc$ ghosts of spin 1
are defined in the usual way (omitting the $^{(1)}$
superscript):\BE\new\BA{l}b(z)=\sum_{n\in{\open Z}}b_nz^{-n-1}~,\qquad
c(z)=\sum_{n\in{\open Z}}c_nz^{-n}~,\\ \{b_n,c_m\}=\Kr{m+n}~,\qquad
b_{\geq0}\ket0_{{\myrm{gh}}}=c_{>0}\ket0_{{\myrm{gh}}}=0~.\EA\EE Therefore,
\BE Q^{(1)}(z)=b(z)\label{Q}\EE is now of spin 1, and we also get the spin-2
odd generator\BE\cG^{(1)}(z)=2c(z)\cT^{(1)}(z)-2\sqrt{{3-\ctop\over3}}\,\d
c(z)\!\cdot\!I(z)+{\ctop\over3}\,\d^2c(z)\,.\label{G}\EE

Henceforth, we will omit the $^{(1)}$ superscript, as we are going to deal
exclusively with this `spin-1' construction for the twisted algebra
\req{topalgebra}. The other twisting of the $N\!=\!2$ algebra
\req{N2algebra}, leading to spin-2 $bc$ ghosts, can be considered similarly
\cite{[GS3]}.

\medskip

The realisation \req{construction1}--\req{G} of the topological conformal
algebra \req{topalgebra} can be applied to its singular states. Demanding
that a singular vector $\ket{\Upsilon}$ decouple amounts to the vanishing of
all the correlators $\L<\Upsilon(z_a)\prod_{b\neq a}\Psi_b(z_b)\R>$ where
$\Psi_b$ are chiral primary fields (in the twisted version, which will be
understood in the following). Now, having built up a singular vector
$\ket{\Upsilon}$ at level\/ $l$, we can substitute into it the above
expressions for the topological generators in terms of the `costituent'
matter, Liouville, and ghosts. Note that $\Psi$ will be a chiral primary if
we set $\ket{\Psi}\!=\!\ket{\myrm{matter}}\otimes\ket{\myrm{Liouville}}
\otimes\ket0_{{\myrm{gh}}}$, with
$\ket{\myrm{matter}}\otimes\ket{\myrm{Liouville}}$ being the matter primary
state dressed by the Liouville according to a certain prescription\footnote{
The Liouville charges $n_b$
must be related to the topological $U(1)$ charges as
\BE\htop_b=\sqrt{{3-\ctop\over3}}n_b\label{othern}\EE}. Thus a reduction to
the matter$\otimes$Liouville theory is well-defined (and a subsequent
reduction to the matter theory alone would reproduce the standard minimal
singular states). The singular states thus obtained can also be constructed
directly, as null states satisfying the `Kontsevich--Miwa' dressing condition
\cite{Solving,[GS3]}. Decoupling equations associated with these singular
vectors are implemented by order-$l$ differential operators (where $l$ is the
level):\BE\cO^{(l)}_a\Bigl<\Psi(z_a)\prod_{b\neq a}\Psi_b(z_b)\Bigr>=0\EE and
such operators $\cO^{(l)}_a$ factorise (completely, or up to a certain
obstruction) through the combination $\sum_{n\geq-1}z_a^{-n-2}\sL_n$ of the
Virasoro generators\BE\new\BA{rcl}\sL_{p>0}
&=&\half\sum^{p-1}_{s=1}{\d^{2}\over\d t_{p-s}\d t_s}+\sum_{s\geq 1}st_s
{\d\over\d t_{p+s}}+\half\sQ(p+1) {\d\over\d t_p}\,,\\ \sL_0&=&\sum_{s\geq
1}st_s {\d\over\d t_s}\,,\qquad{\sf L}_{-1}~=~\sum_{s\geq
1}(s+1)t_{s+1}{\d\over\d t_s}\,,\EA\label{Lontau}\EE which are written down
in terms of the time parameters introduced via the Kontsevich--Miwa transform
\cite{Solving}\BE t_r={1\over r}\sum_b n_bz^{-r}_b,\quad
r\geq1\label{Miwatransform}\EE where the $n_b$ are the Liouville charges of
the fields $\Psi_b$. The \bc\ $\sQ$ of these Virasoro generators,
$\sQ={2n_a\over l-1}-{l-1\over n_a}$, turns out to coincide with the matter
\bc\ $Q\!=\!\sqrt{(1-d)/3}$. A full factorisation occurs only for the
$(l,1)$- or $(1,l)$-type singular vectors (see the next section), while for
the type-$(r,s)$ vectors with both numbers different from 1, there is an
obstruction to the full factorisation \cite{[GS3]}. The `invariant' mechanism
(in terms of the structure of topological singular vectors) behind these
factorisation properties is unclear, but it is very likely related to the
hidden $sl(2)$ structure that we study below.

\section{Topological singular vectors}\lvm Two integers $r\equiv2j_1+1$ and
$s\equiv2j_2+1$ characterise a topological singular vector $\ket\Upsilon$ in
the following way: $\ket{\Upsilon}$ is the singular vector on level $l=rs$
upon the chiral primary state $\ket\htop$ whose topological $U(1)$ charge
$\htop$ is related to the \tcc\ $\ctop$ by
\BE\htop={\ctop-3\over3}j_1+2j_2\label{therelation}\EE (in the language of
\cite{[BFK]}, this is the {\bf A} series with zero relative charge; see also
\cite{[Kir]}).

For example, at level 4, we thus get three possibilities according to the
values of $\{j_1,j_2\}$:

\BE\left\{\new\BA{ll}\{\frac{3}{2},0\},& \htop=\half(\ctop-3),\\
\{\frac{1}{2},\half\},&\htop=\sixth(\ctop+3),\\
\{0,\frac{1}{2}\},&\htop=3\,.\EA\right.\label{cases}\EE The corresponding
singular vectors follow from a `prototype' state written out in
eq.~\req{proto} upon substituting into it one of these formulas for the
topological $U(1)$ charge $\htop$.

Thus in the first case, that of $\{j_1,j_2\}\!=\!\{\frac{3}{2},0\}$ and
$\htop\!=\!(\ctop-3)/2$, eq.~\req{proto} takes the form
\BE-\frac{1}{12}(\ctop-1)(\ctop-3)^4
\ket{\Upsilon^{(4)}_{\{{3\over2},0\}}}\,,\EE where the `normalised' singular
vector $\ket{\Upsilon^{(4)}_{\{{3\over2},0\}}}$ (whose normalisation, only
possible for $\ctop\neq3$, has been chosen in anticipation of the coincidence
with the corresponding MFF state) equals {\tenpt
\BE\new\BA{l}\ket{\Upsilon^{(4)}_{\{{3\over2},0\}}}=
{108\over(\ctop-3)^2}\biggl( {72(5-2\ctop)\over(\ctop-3)^2}\cH_{-4}+
{2(18+12\ctop-11\ctop^2+\ctop^3)\over(\ctop-3)^2}\cL_{-4}+
{3(2\ctop-13)\over\ctop-3}(\cG_{-3}\cQ_{-1}-2\cL_{-3}\cH_{-1})\\ {}-
{2(51-13\ctop+2\ctop^2)\over(\ctop-3)^2}\cH_{-3}\cL_{-1}
+{6(1-\ctop)\over\ctop-3}\cH_{-2}\cL_{-2}+
{4(21-16\ctop+2\ctop^2)\over(\ctop-3)^2}\cL_{-3}\cL_{-1}+3\cL_{-2}^2\\ {}+
{(225+13\ctop-2\ctop^2)\over(\ctop-3)^2}\cQ_{-3}\cG_{-1}-
{3(33-16\ctop+\ctop^2)\over(\ctop-3)^2}\cQ_{-2}\cG_{-2}
+{36\over3-\ctop}\cG_{-2}\cH_{-1}\cQ_{-1}+
{18(\ctop-2)\over(\ctop-3)^2}\cG_{-2}\cL_{-1}\cQ_{-1}\\ {}+
{6(3\ctop-2)\over(\ctop-3)^2}(\cH_{-2}\cQ_{-1}\cG_{-1}+
2\cH_{-2}\cH_{-1}\cL_{-1})+ {4(3-5\ctop)\over(\ctop-3)^2}\cH_{-2}\cL_{-1}^2+
{36\over\ctop-3}\cL_{-2}\cH_{-1}^2+
{60\over3-\ctop}\cL_{-2}\cH_{-1}\cL_{-1}\\ {}+
{20\over\ctop-3}\cL_{-2}\cL_{-1}^2+
{6(9-2\ctop)\over(\ctop-3)^2}\cL_{-2}\cQ_{-1}\cG_{-1}+
{18(\ctop-14)\over(\ctop-3)^2}\cQ_{-2}\cH_{-1}\cG_{-1}+
{2(72-5\ctop)\over(\ctop-3)^2}\cQ_{-2}\cL_{-1}\cG_{-1}\\ {}-
{72\over(\ctop-3)^2}\cH_{-1}^3\cL_{-1}+
{132\over(\ctop-3)^2}\cH_{-1}^2\cL_{-1}^2-
{108\over(\ctop-3)^2}\cH_{-1}^2\cQ_{-1}\cG_{-1}-
{72\over(\ctop-3)^2}\cH_{-1}\cL_{-1}^3\\ {}+
{132\over(\ctop-3)^2}\cH_{-1}\cL_{-1}\cQ_{-1}\cG_{-1}+
{12\over(\ctop-3)^2}\cL_{-1}^4-
{36\over(\ctop-3)^2}\cL_{-1}^2\cQ_{-1}\cG_{-1}
\biggr)\ket{(\ctop-3)/2}\label{vect1c}\EA\EE} Next, by substituting into
\req{proto} $\{j_1,j_2\}\!=\!\{\frac{1}{2},\frac{1}{2}\}$,
$\htop\!=\!(\ctop+3)/6$, we bring \req{proto} to the form
${-5\over108}(\ctop-3)^6\ket{\Upsilon^{(4)}_{\{{1\over2},{1\over2}\}}}$ with
the `normalised' topological singular state being given by {\tenpt\BE\new
\BA{l}\ket{\Upsilon^{(4)}_{\{{1\over2},{1\over2}\}}}={36\over(\ctop-3)^4}
\biggl({72(-45+6\ctop-5\ctop^2)\over(\ctop-3)^2}\cH_{-4}-
{12(135-9\ctop+15\ctop^2-\ctop^3)\over(\ctop-3)^2}\cL_{-4}\\{}
+{6(9-30\ctop+\ctop^2)\over\ctop-3}\cH_{-3}\cL_{-1}
-{(27+6\ctop-\ctop^2)^2\over (\ctop-3)^2}\cL_{-2}^2
-{6(135+27\ctop-3\ctop^2+\ctop^3)\over(\ctop-3)^2}\cH_{-2}\cL_{-2}\\{}
+{3(135+45\ctop-3\ctop^2-\ctop^3)\over(\ctop-3)^2}(2\cL_{-3}\cH_{-1}-
\cG_{-3}\cQ_{-1})
-{12(81+9\ctop+3\ctop^2-\ctop^3)\over(\ctop-3)^2}\cL_{-3}\cL_{-1}\\{}
-{3(135-315\ctop-3\ctop^2-\ctop^3)\over(\ctop-3)^2}\cQ_{-3}\cG_{-1}+
{3(135-63\ctop+33\ctop^2-\ctop^3)\over(\ctop-3)^2}\cQ_{-2}\cG_{-2}\\{}
+{36(3+\ctop)^2\over(\ctop-3)^2}(\cL_{-2}\cH_{-1}^2-\cG_{-2}\cH_{-1}\cQ_{-1})
+{36(18+3\ctop+\ctop^2)\over(\ctop-3)^2}(\cH_{-2}\cQ_{-1}\cG_{-1}+
2\cH_{-2}\cH_{-1}\cL_{-1})\\{}
{}+{72(9\!-\!3\ctop\!+\!\ctop^2)\over(\ctop-3)^2}\cG_{-2}\cL_{-1}\cQ_{-1}
+{108(3+\ctop)\over3-\ctop}\cH_{-2}\cL_{-1}^2
-{12(135+27\ctop-3\ctop^2+\ctop^3)\over(\ctop-3)^2}
\cL_{-2}\cH_{-1}\cL_{-1}\\{}
{}+{12(45-6\ctop+\ctop^2)\over\ctop-3}\cL_{-2}\cL_{-1}^2-
{6(27+63\ctop-15\ctop^2+\ctop^3)\over(\ctop-3)^2}\cL_{-2}\cQ_{-1}\cG_{-1}-
{648\ctop\over(\ctop-3)^2}\cQ_{-2}\cH_{-1}\cG_{-1}\\{}
-{36(9-12\ctop-\ctop^2)\over(\ctop-3)^2}(\cH_{-1}^2\cL_{-1}^2+
\cH_{-1}\cL_{-1}\cQ_{-1}\cG_{-1})
+{36(3+\ctop)\over3-\ctop}(\cL_{-1}^2\cQ_{-1}\cG_{-1}+2\cH_{-1}\cL_{-1}^3)\\{}
-{36(9-6\ctop-\ctop^2)\over(\ctop-3)^2}\cQ_{-2}\cL_{-1}\cG_{-1}
-{108(3+\ctop)\over(\ctop-3)^2}(2\cH_{-1}^3\cL_{-1}+
3\cH_{-1}^2\cQ_{-1}\cG_{-1})+36\cL_{-1}^4\biggr)\ket{(\ctop+3)/6}
\label{vect2c}\EA\EE
}
Finally, in the case $\{j_1,j_2\}\!=\!\{0,\frac{3}{2}\}$, $\htop\!=\!3$,
eq.~\req{proto} becomes $${(\ctop+15)(3-\ctop)^7\over7776}\cdot
\ket{\Upsilon^{(4)}_{\{0,{3\over2}\}}}$$ where the third level-4 singular
state equals {\tenpt\BE\new\BA{l}
\ket{\Upsilon^{(4)}_{\{0,{3\over2}\}}}={1296\over(\ctop-3)^4}\biggl(
{216(15+7\ctop)\over(3-\ctop)^3}\cH_{-4}+
{36(\ctop-15)\over(\ctop-3)^2}\cL_{-4}+
{270\over(\ctop-3)^2}\cG_{-3}\cQ_{-1}\\{}+
{12(207+\ctop^2)\over(3-\ctop)^3}\cH_{-3}\cL_{-1}+
{216(15+\ctop)\over(3-\ctop)^3}\cH_{-2}\cL_{-2}-
{540\over(\ctop-3)^2}\cL_{-3}\cH_{-1}+
{12(5\ctop-6)\over(\ctop-3)^2}\cL_{-3}\cL_{-1}\\{}+
{324\over(\ctop-3)^2}\cL_{-2}^2+
{30(9+\ctop)\over(\ctop-3)^2}\cQ_{-3}\cG_{-1}+
{270\over(\ctop-3)^2}\cQ_{-2}\cG_{-2}+
{3888\over(3-\ctop)^3}\cG_{-2}\cH_{-1}\cQ_{-1}\\{}+
{378\over(\ctop-3)^2}\cG_{-2}\cL_{-1}\cQ_{-1}
+{18(87+7\ctop)\over(\ctop-3)^3}(\cH_{-2}\cQ_{-1}\cG_{-1}+
2\cH_{-2}\cH_{-1}\cL_{-1})
-{24(12+\ctop)\over(\ctop-3)^2}\cH_{-2}\cL_{-1}^2\\{}
+{3888\over(\ctop-3)^3}\cL_{-2}\cH_{-1}^2-
{1080\over(\ctop-3)^2}\cL_{-2}\cH_{-1}\cL_{-1} {}+
{60\over\ctop-3}\cL_{-2}\cL_{-1}^2-
{162\over(\ctop-3)^2}\cL_{-2}\cQ_{-1}\cG_{-1}\\{}-
{270\over(\ctop-3)^2}\cQ_{-2}\cH_{-1}\cG_{-1}+
{6(4\ctop-9)\over(\ctop-3)^2}\cQ_{-2}\cL_{-1}\cG_{-1}+
{1296\over(3-\ctop)^3}\cH_{-1}^3\cL_{-1}
+{1944\over(3-\ctop)^3}\cH_{-1}^2\cQ_{-1}\cG_{-1}\\{}+{36\over3-\ctop}
\cH_{-1}\cL_{-1}^3
+{396\over(\ctop-3)^2}(\cH_{-1}^2\cL_{-1}^2+\cH_{-1}\cL_{-1}\cQ_{-1}\cG_{-1})
+\cL_{-1}^4+{18\over3-\ctop}\cL_{-1}^2\cQ_{-1}\cG_{-1}\biggr)\ket3
\label{vect3c}\EA\EE} Let us note that pairwise intersections between the
three classes of singular vectors occur precisely for those values of $\htop$
and $\ctop$ that satisfy a chosen pair of conditions \req{cases}
simultaneously.

The case when \tcc\ $\ctop\!=\!3$ is discussed in section~8.

\smallskip

Continuing with the examples, let us go down to level 3. Here, for
$j_1\!=\!1$, $j_2\!=\!0$, $\htop\!=\!(\ctop-3)/3$, the `normalised'
topological singular vector reads
\BE\new\BA{rcl}\ket{\Upsilon^{(3)}_{\{1,0\}}}&=& {12\over(\ctop-3)^2}\Bigl(
2\ctop\cL_{-3}+ {6\ctop\over3-\ctop}\cH_{-2}\cL_{-1}-12\cH_{-1}\cL_{-2}+
12\cL_{-2}\cL_{-1}\\ {}&{}&{}+{3(15-\ctop)\over\ctop-3}\cQ_{-2}\cG_{-1}-
6\cQ_{-1}\cG_{-2}+{36\over\ctop-3}\cH_{-1}^2\cL_{-1}+
{54\over3-\ctop}\cH_{-1}\cL_{-1}^2\\ {}&{}&{}+
{36\over\ctop-3}\cH_{-1}\cQ_{-1}\cG_{-1}+{18\over\ctop-3}\cL_{-1}^3+
{27\over3-\ctop}\cL_{-1}\cQ_{-1}\cG_{-1}\Bigr)\ket{(\ctop-3)/3}
\EA\label{T310}\EE Similarly, at $j_1\!=\!0$, $j_2\!=\!1$, $\htop\!=\!2$ we
find another singular vector\BE\new\BA{rcl}\ket{\Upsilon^{(3)}_{\{0,1\}}}&=&
{216\over(\ctop-3)^3}\biggl( {12(9+\ctop)\over(\ctop-3)^2}\cL_{-3}-
{6(9+\ctop)\over(\ctop-3)^2}\cH_{-2}\cL_{-1}-
{144\over(\ctop-3)^2}\cH_{-1}\cL_{-2}\\ {}&{}&{}+
{24\over\ctop-3}\cL_{-2}\cL_{-1}+{6\over\ctop-3}\cQ_{-2}\cG_{-1}-
{72\over(\ctop-3)^2}\cQ_{-1}\cG_{-2}+
{72\over(\ctop-3)^2}\cH_{-1}^2\cL_{-1}\\ {}&{}&{}+
{18\over3-\ctop}\cH_{-1}\cL_{-1}^2+
{72\over(\ctop-3)^2}\cH_{-1}\cQ_{-1}\cG_{-1}+\cL_{-1}^3+
{9\over3-\ctop}\cL_{-1}\cQ_{-1}\cG_{-1}\biggr)\ket{2}\EA\label{T301}\EE

\smallskip

The situation is yet simpler at level 2. There exist two topological singular
states, one at $\htop\!=\!(\ctop-3)/6$, which reads
\BE\ket{\Upsilon^{(2)}_{\{{1\over2},0\}}}={6\over\ctop-3}\Bigl(\cL_{-2}+
{6\over3-\ctop}\cH_{-1}\cL_{-1}+{6\over\ctop-3}\cL_{-1}^2+
{3\over3-\ctop}\cQ_{-1}\cG_{-1}\Bigr)\ket{(\ctop-3)/6}\label{1top2}\EE and
the other at $\htop\!=\!1$,\BE\ket{\Upsilon^{(2)}_{\{0,{1\over2}\}}}=
{36\over(\ctop-3)^2}\Bigl({6\over\ctop-3}\cL_{-2}-
{6\over\ctop-3}\cH_{-1}\cL_{-1}+\cL_{-1}^2-{3\over\ctop-3}\cQ_{-1}\cG_{-1}
\Bigr)\ket{1}\label{2top2}\EE (these two coincide at $\ctop\!=\!9$,
$\htop\!=\!1$). This completes our construction of lower-level topological
singular states.

\section{$sl(2)_k$ and matter coupled to gravity}\lvm The above singular
vectors can be `split' according to the decomposition
$\frt=\frm\oplus\frl\oplus[bc]$ of the topological algebra, where $\frm$,
$\frl$ and $[bc]$ are the mater, (`mirror') Liouville, and ghost theories
respectively (see section~2). By further suppressing ghosts, we get those
singular states in the matter + Liouville theory which satisfy the
`Kontsevich--Miwa' conditions, and which can be arrived at independently of
the topological considerations \cite{[GS3]}. Dropping down the ghosts is a
meaningful procedure, since ghost-independent field operators represent
chiral primary fields, and thus only the ghost-independent part of a given
singular vector contributes to the decoupling equations for correlators
comprising chiral primary fields. It is to these decoupling equations that
the Kontsevich--Miwa transform can be applied, showing complete factorisation
for the $(l,1)$ or $(1,l)$ states.

Another, and not unrelated, approach allowing us to investigate the structure
of topological singular vectors is provided by revealing a hidden $sl(2)$
Ka\v{c}--Moody algebra. This can be done using the (twisted) Kazama--Suzuki
model \cite{[KS]} $\widetilde{sl}(2)_k\oplus u(1)/u(1)$, where
$\widetilde{sl}(2)_k$ denotes the level-$k$ algebra\BE\new\BA{rcl}
J^0(z)J^\pm(w)&=&{}\pm{J^\pm\over z-w}\,,\qquad
J^0(z)J^0(w)~=~{}{k/2\over(z-w)^2}\,,\\
J^+(z)J^-(w)&=&{}-{k\over(z-w)^2}-{2J^0\over z-w}\,,\EA\label{sl2algebra}\EE
with the twisted Sugawara \emt\BE\widetilde{T^{{\myrm{S}}}}={1\over
k+2}\L(J^0J^0-\half(J^+J^-+J^-J^+)\R)+\d J^0\,.\label{twistedS}\EE

The Kazama--Suzuki `numerator' $u(1)$ algebra fermionises into a couple of
spin-1 ghosts, denoted in the following as $BC$ (cf.~\cite{[MV]}), which
allows us to build up the topological algebra generators in the standard way
\cite{[Ey],[EHy],[Le],[NS]}: The odd generators $\cQ$ and $\cG$ are given by
\BE\cQ=\sqrt{{2\over k+2}}\,BJ^+\,,\qquad\cG=-\sqrt{{2\over
k+2}}\,CJ^-\,,\label{QGsl}\EE while the topological $U(1)$ current and the
\emt\ take the form\BE\cH=-{k\over k+2}BC-{2\over k+2}J^0\label{Hsl}\,,\EE
\BE\cT={}-{1\over k+2}(J^+J^-)+{k\over k+2}\d B\!\cdot\! C+{2\over
k+2}BCJ^0\,.\label{Tsl}\EE Generators \req{QGsl}, \req{Hsl}, and \req{Tsl}
close to the algebra \req{topalgebra} with \tcc\BE\ctop={3k\over
k+2}\,.\label{ctopk}\EE

We thus have a mapping\BE\frt\to\cU(sl(2)_k\oplus
u(1))\,,\label{frttosl(2)}\EE where $\frt$ is the topological algebra and
$\cU$ denotes the universal enveloping. On the other hand, as we have
discussed above, it is possible to view the topological algebra $\frt$ as
$\frt\!=\!\frm\oplus\frl\oplus[bc]$.
We have short exact sequences (omitting $\cU$)\BE\BA{c} 0\\ \downarrow\\
\frm\oplus\frl\oplus[bc]\\ \downarrow\\ 0\to sl(2)\to\phantom{l}\cA\to
u(1)_{BC}\to0\\ \downarrow\\ u(1)_v\\ \downarrow\\
0\EA\label{shortdiagram}\EE where $u(1)_v$ is the Kazama--Suzuki denominator
$u(1)$ algebra, which is generated by the current\BE\d v=\sqrt{{2\over
k+2}}\,(J^0-BC)\label{vscalar}\EE that decouples from the generators
\req{QGsl}, \req{Hsl}, \req{Tsl}. The vertical sequence splits as well as
the horizontal one does, which gives us two ways to describe the algebra
$\cA$. First of all, this algebra is just $sl(2)\oplus u(1)$, but at the same
time it is $\frm\oplus\frl\oplus[bc]\oplus u(1)_v$, and in these latter terms
the splitting of the horizontal sequence is accomplished by representing the
`Kazama--Suzuki' ghosts as\BE B=be^{-\sqrt{{2\over k+2}}(v-\phi)}\,,\quad
C=ce^{\sqrt{{2\over k+2}}(v-\phi)}\,.\label{BCdecouple}\EE Also, the
embedding $sl(2)_k\hookrightarrow\frm\oplus\frl\oplus[bc]\oplus u(1)_v$ is
given by the following explicit
formulae:\BE\new\BA{rcl}J^+&=&e^{\sqrt{{2\over k+2}}(v-\phi)}\,,\qquad
J^0~{}={}~-i+\sqrt{{2\over k+2}}\,I+{k\over\sqrt{2(k+2)}}\d v\,,\\
J^-&=&\bigl\{ -(k+2)(T+T_{{\myrm{L}}})+i^2 -(k+1)\d i-\sqrt{2(k+2)}\,Ii
\bigr\}e^{\sqrt{{2\over k+2}}(\phi-v)}\EA\label{JJJtop}\EE (recall that
$i=-bc$ is the ghost current, $\d\phi\!=\!I$ is the Liouville current, and
$T$ and $T_{{\myrm{L}}}$ are the matter and Liouville \emt s respectively; it
should also be kept in mind that we are dealing with the `mirror' version in
which the Liouville \emt\ has the form\BE
T_{{\myrm{L}}}=-\half\d\phi\d\phi-\half Q\d^2\phi\,,\EE with the \bc\
$Q=\sqrt{(1-d)/3}$\,). A remarkable fact (which was to be expected, though)
is that upon dropping the ghosts, the current $J^-$ reduces to the matter and
Liouville \emt s dressed with an appropriate vertex operator of dimension
zero\footnote{No hamiltonian reduction is being performed!}.

Another remarkable feature is that the matter only enters in
eqs.~\req{JJJtop} (as well as in the construction of the topological algebra
from section~2) through its Virasoro generators and therefore can be
arbitrary. There is in \req{JJJtop} one field more than necessary for
representing three $sl(2)$ currents, and the construction can be
`straightened up' as shown in the next section.

\medskip

Corresponding to the diagram \req{shortdiagram}, we have the following
decomposition of the energy-momentum tensors:\BE
\widetilde{T^{{\myrm{S}}}}-B\d C=\cT+\half\d v\d v
+{k+1\over\sqrt{2(k+2)}}\d^2v\,,\label{decomp}\EE where $\cT$ is the
topological \emt. The realisation of the topological theory as matter dressed
with `mirror' gravity allows us to read the formula \req{decomp} as a
representation for the twisted $sl(2)_k$ theory together with the $BC$ ghosts
as a sum of {\it two\/} minimal models $M_{k+2,1}$, a system of
weight-$(1,0)$ ghosts, and a Liouville scalar whose \bc\ coincides with that
associated to the $M_{k+2,1}$ minimal
theory:\BE\widetilde{sl}(2)_k\oplus[BC]=
\underbrace{M_{k+2,1}^{{\myrm{matter}}}\oplus{\myrm{Liouville}}\oplus[bc]}
\oplus M_{k+2,1}^v\label{twoMdecomp}\EE The $M_{k+2,1}^v$ theory `absorbs the
\cc' (leaving 0 to the topological model) according to\BE\biggl({3k\over
k+2}-6k\biggr)-2=0+\biggl(1-3{2(k+1)^2 \over k+2}\biggr)\,.\EE

\medskip

Next, we would like to represent {\it states\/} in two ways, as
\BE\ket{~}_{sl(2)}\otimes\ket{~}_{BC}=\ket\htop\otimes\ket{~}_{v}
\label{tensor}\EE with $\ket\htop$ being a topological state. This is indeed
possible for {\sl primary\/} states, as
\BE\ket{\{j_1,j_2\}}\otimes\ket0_{BC}=
\ket{\htop_j}\otimes\ket{V_j}\label{states}\EE where $\ket{0}_{BC}$ is the
ghost vacuum:\BE B_{\geq1}\ket0=C_{\geq0}\ket0=0\,,\label{BCvac}\EE and
$\htop_j$ is to be expressed through $\{j_1,j_2\}$ via eq.~\req{therelation}.
{}From \req{Hsl} and the highest-weight conditions we find
\BE\htop=\htop_j=-{2j\over k+2}\,.\label{htopk}\EE Combined with \req{ctopk},
eq.~\req{htopk} allows us to read the equality~\req{therelation} as the
parametrisation \req{jformula} for the highest weight of the $sl(2)$ affine
algebra. Note also that the \bc\ $Q\!=\!\sqrt{(1-d)/3}$ evaluates from
\req{ctopk} and \req{d(c)} as $Q\!=\!{\sqrt{2}(k+1)/\sqrt{k+2}}$.

Further, to find out what the $V_j$ operator is, consider the balance of
dimensions. Since the twisted $N\!=\!2$ dimension of $\ket\htop$ is zero, the
r\^ole of the dressing by $V_j\!=\!\exp\rho_jv$ is to provide the state
$\ket{j}_{sl(2)}$ with the correct (twisted Sugawara) dimension
\BE\widetilde{\delta}[j]={j(j+1)\over k+2}-j\,.\label{twisteddim}\EE Together
with $\ket{~}_{BC}\!=\!\ket0$ this determines $\rho_j$ as
\BE\rho_j=j\sqrt{{2\over k+2}}\qquad {\myrm{or}}\qquad(k+1-j)\sqrt{{2\over
k+2}}\,,\label{vexponents}\EE The first of these values (to be used in
eq.~\req{state} below) coincides with minus the Liouville $U(1)$ charge,
which is determined from \req{therelation} and \req{othern} as
$n_a=-j\sqrt{{2\over k+2}}$.

\section{An `invariant' Wakimoto construction and hamiltonian
reduction\label{sec:aside}}\lvm The construction \req{JJJtop} expresses the
$sl(2)$ currents in terms of three independent scalar fields {\it and\/} a
central-charge-$d$ matter. As in the construction
\req{construction1}--\req{G} for the topological conformal algebra, matter
need not be bosonised, since it enters only through its Virasoro generators.
Anyway, the total of `almost' four bosonic currents is redundant for
representing three $sl(2)$ currents. To remove this redundancy from
eqs.~\req{JJJtop}, we have to find a combination of the $bc$ ghost,
Liouville, and $\d v$- currents that decouples from the $sl(2)$ algebra.
According to \req{shortdiagram} and \req{BCdecouple}, this is the $BC$ ghost
current\BE\d F=i+\sqrt{{2\over k+2}}(\d v-\d\phi)\,.\EE Two other independent
fields are introduced along with $\d F$ via the appropriate mixing,
as\BE\new\BA{rcl} \d\chi&=&\sqrt{{k+2\over k}}\,\d\phi-\sqrt{{2\over
k}}\,i\,,\\ \d\psi&=&\sqrt{{k\over k+2}}\,\d v+{2\over\sqrt{k(k+2)}}\,\d\phi
-\sqrt{{2\over k}}\,i\EA\EE (with signatures $-$ and $+$ respectively, cf.
eq.~\req{signatures}). In terms of these, the $sl(2)$ currents \req{JJJtop}
take the form\BE\new\BA{rcl} J^+&=&e^{\sqrt{2\over k}(\psi-\chi)},\qquad
J^0~=~\sqrt{k\over2}\,\d\psi\,,\\ J^-&=&\Bigl[-(k+2)T+
k\Bigl(\half\d\chi\d\chi+{k+1\over\sqrt{2k}}\,\d^2\chi\Bigl)\Bigl]
e^{-\sqrt{2\over k}(\psi-\chi)}\EA\label{JJJpara}\EE where, as before, $T$ is
the matter \emt\ with \cc\ evaluated from \req{d(c)} and \req{ctopk} as\BE
d=13-6(k+2)-{6\over k+2}\,.\EE Thus an arbitrary matter theory with \cc\
$d\leq1$ or $d\geq25$ can be dressed so as to make up the $sl(2)$
Ka\v{c}--Moody algebra. The matter need not be specified any further beyond
its Virasoro algerba.

\font\tenmathbold=cmmib10 \font\twlmathbold=cmmib10 scaled\magstep1
\def\bvarphi{\mbox{\tenmathbold\char'047}}
\def\bphi{\mbox{\twlmathbold\char'036}} However, {\it if\/} one bosonises the
matter through a scalar with the appropriate \bc, then, by a simple exercise
in linear algebra, one can explicitly map the formulae \req{JJJpara} onto the
more standard Wakimoto bosonisation \cite{[W],[Dko],[GMMOS]}: a (hyperbolic)
rotation in the space of the currents allows one to identify a weight-1
$\beta\gamma$ system, bosonised according to the recipe \cite{[MaSo]}
$\beta={\bf b}e^{-\bvarphi}$, $\gamma=-\d{\bf c}e^{\bvarphi}$, and an
independent scalar $\bphi$. In this way, the currents take the standard
Wakimoto form\BE\new\BA{rcl} J^+&=&\beta\,,\qquad
J^0~=~\beta\gamma+\sqrt{{k+2\over2}}\,\d\bphi\,,\\
J^-&=&\beta\gamma^2+\sqrt{2(k+2)}\,\d\bphi\!\cdot\!\gamma+k\d\gamma\,.
\EA\label{slbos}\EE We do not present the details, partly because they have
already appeared in \cite{[OS]}, and partly because we prefer the form
\req{JJJpara} as being more `intrinsic' (i.e. not requiring a bosonisation
of the matter theory).  In fact, once the bosonisation is allowed, the entire
set of {\sl four\/} fields from eq.~\req{JJJtop} can be organised to make up
a Wakimoto-bosonised $osp(2,1)$ Ka\v{c}--Moody algebra (cf.~\cite{[BO]}).
Also, the matter${}+\d\chi$ theory from \req{JJJpara} is made equivalent to
$\oZ_k$-parafermions, as it should be after `suppressing' the $J^0$ current
in the $sl(2)_k$ WZW model.

\medskip

As an aside, note that equations~\req{JJJpara} suggest themselves
for the hamiltonian reduction (cf. \cite{[BO1],[GMM-h]}). This
amounts to projecting out the $\psi$ and $\chi$ scalars, and we thus
get from the theory with the \emt\ \req{decomp} precisely the {\sl matter\/}
$M_{k+2,1}$ theory (see \req{twoMdecomp}), plus the $BC$ ghosts. The ghosts
are not interesting in this context, since they remain a direct summand and
are not involved in the reduction. Thus, the matter theory is recovered by
performing the hamiltonian reduction of the $\widetilde{sl}(2)_k$ theory,
which gives the matter an `invariant' meaning in the context of decomposing a
given topological conformal theory as $\frt=\frm\oplus\frl\oplus[bc]$.

\section{The MFF vectors}\lvm We will use the mapping \req{frttosl(2)} for an
explicit evaluation of the topological singular states. Namely, we are going
to show that singular vectors of the topological conformal algebra coincide
with those of the $sl(2)$ Ka\v{c}--Moody algebra.

These latter are well known: singular vectors of $sl(2)_k$ are labelled by
two integers $r$ and $s$, and can be written in the MFF form \cite{[MFF]} as
\BE\new\BA{rcl}\ket{{\myrm{MFF}}{rs}}&=&
(J^-_0)^{r+(s-1)(k+2)}(J^+_{-1})^{r+(s-2)(k+2)}(J^-_0)^{r+(s-3)(k+2)}
\ldots\\ {}&{}&{}\times (J^+_{-1})^{r-(s-2)(k+2)}
(J^-_0)^{r-(s-1)(k+2)}\ket{\{\frac{r-1}{2},\frac{s-1}{2}\}}\EA\label{mff}\EE
where $\ket{\{j_1,j_2\}}$ is a highest-weight state:\BE\new\BA{rcl}
J_n^+\ket{\{j_1,j_2\}}&=&0\,,\quad n\geq0\,,\qquad
J_n^-\ket{\{j_1,j_2\}}~{}={}~0\,,\quad n>0\,,\\
J_n^0\ket{\{j_1,j_2\}}&=&0\,,\quad n>0\,,\qquad
J_0^0\ket{\{j_1,j_2\}}~=~j\ket{\{j_1,j_2\}}\,,\EA\label{hw}\EE with\BE
j=j_1-j_2(k+2)\,.\label{jformula}\EE The formula \req{mff} is not so
innocuous as it might appear: it {\it is\/} simple only when $s\!=\!1$ (being
reduced then to $(J^-_0)^r$), while, for instance, writing out the state
$\ket{{\myrm{MFF}}{14}}$ as a {\it polynomial\/} in the currents one finds
{\tenpt\BE\new\BA{l} (k+3)\biggl(6(k+1)(k+2)^2(3+2k)(7+3k)J^-_{-3}+
12k(k+2)(5+2k)(7+3k)J^-_{-2} J^0_{-1}\\{}-4(7+3k)(4-9k-9k^2-2k^3)J^0_{-3}
J^-_0+6(k+2)(7+3k)(2+7k+2k^2)J^0_{-2} J^-_{-1}\\{}-
2(7+3k)(4-9k-4k^2)J^-_{-2} J^+_{-1} J^-_0+3(k+2)(k+4)(7+3k)J^-_{-1} J^-_{-1}
J^+_{-1}\\{}+12(k+2)(5+2k)(7+3k)J^-_{-1} J^0_{-1} J^0_{-1}+
4(7+3k)(2+17k+6k^2)J^0_{-2} J^0_{-1} J^-_0\\{}-(8-3k-2k^2)[2(k+3)^2J^+_{-3}
J^-_0 J^-_0+2(7+3k)J^+_{-2} J^-_{-1} J^-_0]+4(7+3k)(14+5k)J^-_{-1} J^0_{-1}
J^+_{-1} J^-_0\\{}-4(1-13k-5k^2)J^0_{-2} J^+_{-1} J^-_0 J^-_0+
8(5+2k)(7+3k)J^0_{-1} J^0_{-1} J^0_{-1} J^-_0\\{}-8(9-2k-2k^2)J^+_{-2}
J^0_{-1} J^-_0 J^-_0+2(14+5k)J^-_{-1} J^+_{-1} J^+_{-1} J^-_0J^-_0+
4(29+11k)J^0_{-1} J^0_{-1} J^+_{-1}J^-_0 J^-_0\\{}+4(k-2)J^+_{-2}J^+_{-1}
J^-_0 J^-_0 J^-_0+12J^0_{-1} J^+_{-1} J^+_{-1} J^-_0 J^-_0J^-_0\biggr){}+
J^+_{-1} J^+_{-1} J^+_{-1} J^-_0J^-_0 J^-_0 J^-_0
\EA\label{mff14example}\EE}(which is understood to act on the corresponding
highest-weight state), and a similar expression for $\ket{{\myrm{MFF}}{15}}$
contains already as many as 42 different terms. All of them appear as a
result of the repeated use of the following identities:
\BE\new\BA{rcl}(J^-_0)^qJ^0_m&=&qJ^-_m(J^-_0)^{q-1}+J^0_m(J^-_0)^q\,,\\ {}
(J^-_0)^qJ^+_m&=&J^+_m(J^-_0)^q+2qJ^0_m(J^-_0)^{q-1}+
q(q-1)J^-_m(J^-_0)^{q-2}\,,\\{}
J^0_m(J^+_{-1})^q&=&q(J^+_{-1})^{q-1}J^+_{m-1}+(J^+_{-1})^q J^0_m\,,\\{}
J^-_m(J^+_{-1})^q&=&(J^+_{-1})^q J^-_m+2q(J^+_{-1})^{q-1}J^0_{m-1}+
q(q-1)(J^+_{-1})^{q-2} J^+_{m-2}-kq\delta_{m-1}(J^+_{-1})^{q-1}\,.\EA\EE

The MFF states satisfy the highest-weight conditions
\BE\new\BA{rclcl}J_n^+\ket{S}&=&0\,,&{}&n\geq0\,,\\ J_n^0\ket{S}
&=&0\,,&{}&n\geq1\,,\\ J_n^-\ket{S}&=&0\,,&{}&n\geq2\,. \EA\label{sl2hw}\EE
Let us also note that in order to end up with singular states which are
$J^0$-neutral and belong to level $l=rs$, one has to act on an MFF state
$\ket{{\myrm{MFF}}rs}$ with $(J^+_{-1})^r$.

\pagebreak[3] \section{The evaluation} \hbox
to\hsize{\hfill\parbox{.45\hsize}{\begin{flushright} {\small\sf Lay down the
book, and I will allow you half a day to give a probable guess at the grounds
of this procedure.}

\nopagebreak
{{\small\sc Lawrence Sterne,} {\sl Tristram Shandy\/} (1760)}
\end{flushright} }}

\nopagebreak Now we are ready to formulate the main observation: evaluating
the topological singular states $\ket{\Upsilon}$ in the Ka\v{c}--Moody terms
via the `Kazama--Suzuki mapping' \req{frttosl(2)}, we find the following
identifications:

\BE\ket{\Upsilon}\otimes\ket{V}_v=\ket{S}_{sl(2)}\otimes\ket0_{BC}
\label{mapping}\EE

\noindent where {\sl the $BC$ oscillators drop from the RHS and $\ket{S}$ is
an $sl(2)$ singular vector\/}. In this formula, $V$ is a primary
chosen as explained in the end of section~4; in particular, there are no $\d
v$-oscillators, and therefore the topological singular vector can be {\it
identified\/} with the $sl(2)$ one.

The identity between topological singular states and the MFF vectors
\req{mff} has been checked explicitly for levels 2 through 4. As a by-product
of the computation, we obtain a useful form of the MFF vectors that
involves the Sugawara Virasoro generators; these are not present explicitly
in the original formulation \req{mff}, but are quite helpful when writing
down the decoupling equations corresponding to singular states: using the
standard Ward identities for the Virasoro generators, the $sl(2)$ decoupling
equations are then written as {\it differential\/}.

We find the equalities, presented below, between singular vectors of the
algebras \req{topalgebra} and \req{sl2algebra} very convincing, considering
the relative complexity of some of the expressions that follow.  A general
proof might be easier in the direction from $sl(2)$ to the topological
algebra. It is in fact straightforward to see that any $sl(2)$ primary state,
which satisfies eqs.~\req{sl2hw}, would correspond to a {\it topological\/}
primary via eq.~\req{states}. Indeed, using the mode expansions
$\cQ(z)=\sum\cQ_nz^{-n-1}$ and $\cG(z)=\sum\cG_nz^{-n-2}$, we find from
\req{QGsl} and \req{BCvac} that, on our highest-weight state,
$\cQ_{\geq0}\sim0$ and $\cG_{\geq1}\sim0$. Similarly, eqs.~\req{BCvac} and
\req{Hsl} imply $\cH_{\geq1}\sim0$, whence all the topological highest-weight
conditions follow. Further, the relations \req{ctopk} and \req{htopk} allow
us to rewrite the formula \req{jformula} as \req{therelation}, thus
recovering the {\bf A} series with zero `relative charge' from
ref.~\cite{[BFK]}. However, this does not immediately imply that the $sl(2)$
{\sl singular\/} vectors would be those with respect to the topological
conformal algebra, since it is not clear (to the author) how one can
independently characterise those $sl(2)$-descendants that translate into the
topological descendants\footnote{ Even the converse (that a topological
descendant is an $sl(2)$ descendant) may not be true for any vector other
than a singular one, because, when rewriting a topological state by using
eqs.~\req{QGsl}, \req{Hsl}, and \req{Tsl}, it is not at all obvious that the
$BC$ modes would drop out. For instance, even for the expression \req{proto},
which is `very close' to being a singular state, the $BC$ ghosts do not
decouple unless we substitute one of the relations \req{cases}, thus reducing
\req{proto} to a singular vector.}.

Anyway, along with the results on explicit evaluation of topological
singular vectors in the $sl(2)$ terms, this partial argument in favour of the
one-to-one correspondence between the singular states strongly suggests that
the $sl(2)$ Ka\v{c}--Moody algebra and the topological conformal algebra
possess identical singular vectors (when $\ctop\neq3$, see the next section).
Recall also that `the same' ordinary matter theory enters our constructions
for the topological algebra (section~2) and the $sl(2)$ algebra (section~5),
which are the two algebras whose singular states are identical. We thus
arrive at the diagram
\BE\new\BA{rcl}{{\myrm{topological}}\atop{\myrm{singular\ states}}}
&\stackrel{\sim}{\longleftrightarrow}&{sl(2)\atop{\myrm{singular\ states}}}\\
\searrow&{}&\swarrow\\ {}&{{\myrm{matter}}\atop{\myrm{singular\ states}}}&{}
\EA\label{trianglediagram}\EE As noted above, it would be interesting to
understand these relations not just between the singular states, but rather
between the corresponding algebras, in the context of universal string theory
\cite{[BV],[Fof],[IK]}.

\bigskip

In the remaining part of this section, we present the `experimental evidence'
on the identity between the topological and $sl(2)$ singular vectors.  By a
direct (although quite lengthy) calculation we can evaluate that, for
instance, level-4 state \req{vect1c} gives rise, according to
{}~\req{mapping}, to the following $sl(2)$-state:

{\tenpt\BE\new\BA{l} \ket{S^{(4)}_{\{{3\over2},0\}}}{}=(k+2)\Bigl[
3(-112-102k-11k^2+3k^3)J^0_{-4}+3(8+20k+3k^2-k^3)\tLS_{-4}\\{}-
3(6-4k+k^2)J^-_{-3}J^+_{-1}+3(4-k)(k+2)J^-_{-2}J^+_{-2}-
3(96+59k+4k^2)J^0_{-3}J^0_{-1}\\{}+3(k-4)(k+2)(7+2k)J^0_{-3}\tLS_{-1}+
3(-6-7k+k^2)J^0_{-2}J^0_{-2}+(4-k)(11+3k)J^+_{-4}J^-_0\\{}+
3(22+6k-k^2)J^+_{-3}J^-_{-1}+9(k+2)(k+4)\tLS_{-3}J^0_{-1}+
(k+2)(28-4k-3k^2)\tLS_{-3}\tLS_{-1}\\{}+9(k+2)(\tLS_{-2})^2+
9(k+4)J^-_{-2}J^0_{-1}J^+_{-1}+3(4-k)(k+2)J^-_{-2}J^+_{-1}\tLS_{-1}-
18(k+4)J^0_{-2}J^0_{-1}J^0_{-1}\\{}+6(k-4)(k+2)J^0_{-2}J^0_{-1}\tLS_{-1}+
(28+26k+3k^2)J^0_{-2}J^+_{-2}J^-_0+(32+13k)J^+_{-3}J^0_{-1}J^-_0\\{}+
(k+2)(8+5k)J^+_{-3}\tLS_{-1}J^-_0-3(k+2)(k+6)J^+_{-2}\tLS_{-2}J^-_0
+3(k+4)J^+_{-2}J^0_{-1}J^0_{-1}J^-_0\\{} -10(k\!+\!2)^2\tLS_{-2}(\tLS_{-1})^2
+8(k\!+\!2)J^+_{-2}J^0_{-1}\tLS_{-1}J^-_0
+6(k\!+\!2)^2J^+_{-2}(\tLS_{-1})^2J^-_0\!+\! (k\!+\!2)^3(\tLS_{-1})^4
\Bigr]\ket{\{\frac{3}{2},0\}}.\EA\label{S41}\EE}

\noindent where $\tLS_m$ are modes of the twisted Sugawara
\emt~\req{twistedS}. Evaluating these in terms of the currents, we bring
eq.~\req{S41} to the MFF form

\BE\ket{S^{(4)}_{\{{3\over2},0\}}}=J^+_{-1}J^+_{-1}J^+_{-1}J^+_{-1} J^-_{0}
J^-_{0}J^-_{0}J^-_{0}\ket{\{\frac{3}{2},0\}}
=J^+_{-1}J^+_{-1}J^+_{-1}J^+_{-1}\ket{{\myrm{MFF}}41}\label{v1.4}\EE

For the other level-4 topological singular vectors we observe equally
dramatic cancellations. The singular vector \req{vect2c} is mapped in the
same way into
{\tenpt\BE\new\BA{l}\ket{S^{(4)}_{\{{1\over2},{1\over2}\}}}=(k+2)^2\Bigl(
2(k+2)(13+14k+9k^2+2k^3)J^0_{-4}-
(k+2)(5+2k)(2+2k+k^2)\tLS_{-4}\\{}+(1+6k+k^2-k^3)J^-_{-3}J^+_{-1}-
(k+2)(4+2k+k^2)J^-_{-2}J^+_{-2}+{}\\{}
{}+(38+66k+47k^2+14k^3+k^4)J^0_{-3}J^0_{-1}
{}+(k+2)(k+3)(4+2k+k^2)J^0_{-3}\tLS_{-1}\\{}
+(7+8k+6k^2+2k^3)J^0_{-2}J^0_{-2}-(20+18k+7k^2+k^3)J^+_{-4}J^-_0-
(5+4k+3k^2+k^3)J^+_{-3}J^-_{-1}\\{} -(k+1)(k+2)^2(k+4)\tLS_{-3}J^0_{-1}-
(k+2)(6+10k+6k^2+k^3)\tLS_{-3}\tLS_{-1}\\{}+(k+1)^2(k+3)^2(\tLS_{-2})^2-
(k+1)(k+2)(k+4)J^-_{-2}J^0_{-1}J^+_{-1}-
(k+2)(4+2k+k^2)J^-_{-2}J^+_{-1}\tLS_{-1}\\ {}+
2(k+1)(k+2)(k+4)J^0_{-2}J^0_{-1}J^0_{-1}+
2(k+2)(4+2k+k^2)J^0_{-2}J^0_{-1}\tLS_{-1}\\{}+
2(3+3k+k^2)J^0_{-2}J^+_{-2}J^-_0+(2-6k-5k^2-k^3)J^+_{-3}J^0_{-1}J^-_0-
(k+2)(10+8k+k^2)J^+_{-3}\tLS_{-1}J^-_0\\{}+
(2\!+\!10k\!+\!6k^2\!+\!k^3)J^+_{-2}\tLS_{-2}J^-_0-
2(k\!+\!2)(5\!+\!4k\!+\!k^2)\tLS_{-2}(\tLS_{-1})^2+
(k+2)(k+4)J^+_{-2}J^0_{-1}J^0_{-1}J^-_0\\{}
-4(k+2)J^+_{-2}J^0_{-1}\tLS_{-1}J^-_0-
2(k+1)(k+2)J^+_{-2}(\tLS_{-1})^2J^-_0+(k+2)^2(\tLS_{-1})^4
\Bigr)\ket{\{\frac{1}{2},\half\}}\,,\EA\label{S22}\EE} The reader can check
that by inserting now the Sugawara Virasoro generators, this rewrites as
\BE\ket{S^{(4)}_{\{{1\over2},{1\over2}\}}}=J^+_{-1}J^+_{-1}
\ket{{\myrm{MFF}}22}\,.\EE

The third level-4 singular vector~\req{vect3c} (for which
$\{j_1,j_2\}\!=\!\{0,\frac{3}{2}\}$, $j\!=\! {(-6-3k)/2}$) evaluates
similarly as{\tenpt\BE\new\BA{l}\ket{S^{(4)}_{\{0,{3\over2}\}}}=
(k+2)^4\Bigl((\tLS_{-1})^4 -6(k+2)(11+6k)J^0_{-4}-6(k+2)(5+2k)\tLS_{-4}+
9(k+2)(3+4k)J^-_{-3}J^+_{-1}\\{}-21(k+2)J^-_{-2}J^+_{-2}-
9(k+2)(6+7k)J^0_{-3}J^0_{-1}- 21(k+2)^2J^0_{-3}\tLS_{-1}-
36(k+1)(k+2)J^0_{-2}J^0_{-2}\\{}+(3k-5)J^+_{-4}J^-_0{}-
15(k+2)J^+_{-3}J^-_{-1}+9(k+2)^2\tLS_{-3}J^0_{-1}-
(k+2)(4-3k)\tLS_{-3}\tLS_{-1}+9(k+2)^2(\tLS_{-2})^2\\{}+
9(k+2)J^-_{-2}J^0_{-1}J^+_{-1}-21(k+2)J^-_{-2}J^+_{-1}\tLS_{-1}-
18(k+2)J^0_{-2}J^0_{-1}J^0_{-1}+42(k+2)J^0_{-2}J^0_{-1}\tLS_{-1}\\{}-
(31+18k)J^0_{-2}J^+_{-2}J^-_0-(2+3k)J^+_{-3}J^0_{-1}J^-_0-
(k+10)J^+_{-3}\tLS_{-1}J^-_0+9(k+2)J^+_{-2}\tLS_{-2}J^-_0\\{}-
10(k+2)\tLS_{-2}(\tLS_{-1})^2-3J^+_{-2}J^0_{-1}J^0_{-1}J^-_0+
8J^+_{-2}J^0_{-1}\tLS_{-1}J^-_0-6J^+_{-2}(\tLS_{-1})^2J^-_0
\Bigr)\ket{\{0,\frac{3}{2}\}}\EA\EE} which is nothing other but
\BE\ket{S_{\{0,{3\over2}\}}}=J^+_{-1}\ket{{\myrm{MFF}}14}\EE (the polynomial
expression for $\ket{{\myrm{MFF}}14}$ was written out in \req{mff14example}).

\smallskip

The same situation repeats, in simpler terms, at level 3: the topological
singular vector~\req{T310} evaluates in the Ka\v{c}--Moody language as
\BE\new\BA{rcl}\ket{S^{(3)}_{\{1,0\}}}&=& (k+2)\Bigl(
2k\tLS_{-3}-2kJ^+_{-3}J^-_{0}+4J^+_{-2}J^-_{-1}+4(k+2)\tLS_{-2}\tLS_{-1}\\
{}&{}&{}-2J^+_{-2}J^0_{-1}J^-_{0}-
3(k+2)J^+_{-2}\tLS_{-1}J^-_{0}-(k+2)^2(\tLS_{-1})^3\Bigr)\ket{\{1,0\}}\\
{}&=&{} J^+_{-1}J^+_{-1}J^+_{-1}\ket{{\myrm{MFF}}31}\EA\EE The second level-3
singular state~\req{T301} becomes, similarly,
\BE\new\BA{rcl}\ket{S^{(3)}_{\{0,1\}}}&=& (k+2)^3\Bigl(
-2(k+2)(3+2k)\tLS_{-3}-2(3+2k)J^+_{-3}J^-_{0}-4(k+2)J^+_{-2}J^-_{-1}\\
{}&{}&{}+4(k+2)\tLS_{-2}\tLS_{-1}-2J^+_{-2}J^0_{-1}J^-_{0}
+3J^+_{-2}\tLS_{-1}J^-_{0}-(\tLS_{-1})^3\Bigr)\ket{\{0,1\}}\\ {}&=&{}
J^+_{-1}\ket{{\myrm{MFF}}13}\EA\EE

\smallskip

Finally, at level 2, the first of the two topological null vector rewrites
simply as \BE\new\BA{rcl}\ket{S^{(2)}_{\{{1\over2},0\}}}&=& (k+2)\bigl(
-\tLS_{-2}+J^+_{-2}J^-_0+(k+2)(\tLS_{-1})^2\bigr)\ket{\{\half,0\}}\\
{}&=&J^+_{-1}J^+_{-1}\ket{{\myrm{MFF}}21}\EA\label{1S2}\EE while the other
one, as\BE\new\BA{rcl}\ket{S^{(2)}_{\{0,{1\over2}\}}}&=&(k+2)^2\bigl(
-(k+2)\tLS_{-2}-J^+_{-2}J^-_0+(\tLS_{-1})^2\bigr)\ket{\{0,\half\}}\\
{}&=&J^+_{-1}\ket{{\myrm{MFF}}12}\EA\label{2S2}\EE All these demonstrate a
remarkable correspondence with the MFF states.

\section{$k\to\infty$ and alternative free-field constructions}\lvm Up to
now, the normalisation of the {\it topological\/} singular states has been
chosen in anticipation of their coincidence with the MFF states. This
obviously failed at $\ctop=3$, which corresponds to the classical limit
$|k|\to\infty$. It is worth mentioning that the construction of topological
theories in terms of matter dressed with gravity fails at the same value
$\ctop\!=\!3$ (see eq.~\req{d(c)}). However, from the point of view of the
topological algebra as such, there is nothing singular about this value, and
the null states continue smoothly to this point: it suffices to multiply the
above null states by the minimal power of $(\ctop-3)$ that is necessary, and
then set $\ctop\!=\!3$. Or, at level 4, for instance, we can simply put
$\ctop\!=\!3$ in the `prototype' state \req{proto} and insert the appropriate
value of $\htop$ from \req{cases}. This produces three states given in the
Appendix.

\bigskip

It is possible to trace the `classical' degeneration of the Kazama--Suzuki
mapping \req{QGsl}, \req{Hsl} and \req{Tsl} at $|k|\to\infty$. While in the
original representation \req{mff} the limit of very large $k$ is unclear, it
can be evaluated using the `Sugawara' form of the singular states, e.g.
eq.~\req{S41}. This requires rescaling the currents as
$J^{0,\pm}\mapsto\sqrt{k}J^{0,\pm}$, after which we are left, at
$|k|\!=\!\infty$, with a `complex' scalar current
$J^\pm=(\d\varphi,\d\bar\varphi)$ and an independent current $J^0$ that
decouples from the $\ctop\!=\!3$ topological algebra, while the algebra
itself is now constructed from $(\d\varphi,\d\bar\varphi)$ and the $BC$
ghosts (which are not rescaled). The Sugawara \emt\ then reduces to the \emt\
of two bosons, whose \cc\ $c\!=\!2$ is compensated by that of the $BC$
ghosts. The topological \emt, determined from either \req{Tsl} or
\req{decomp}, becomes\BE\cT=-\d\varphi\,\d\bar\varphi+\d B\!\cdot\!C\qquad
(\ctop=3)\,,\label{Tc3}\EE while the other topological generators reduce
to\BE\cH=-BC\,,\quad\cQ=\sqrt{2}\,B\d\varphi\,,\quad
\cG=-\sqrt{2}\,C\d\bar\varphi\qquad (\ctop=3)\,.\label{HQGc3}\EE Thus the
{\it topological\/} singular states at $\ctop\!=\!3$ can be viewed as a
`resolution' of the MFF construction at $k=\infty$.

\medskip

Interestingly, the formulae \req{Tc3} and \req{HQGc3} represent also the
classical limit of Witten's free-field realisation \cite{[W-LG]} of the
topological conformal algebra, derived from the Landau--Ginzburg theory (to
be precise, the formulae given below are obtained from those of
ref.~\cite{[W-LG]} by twisting, cf.~section~2). Recall that the idea behind
the construction of a topological conformal algebra out of matter, Liouville,
and {\sl spin-1\/} ghosts \cite{[GS2],[GS3]} was that matter and the
(`mirror'!) Liouville give the total \cc\ of \ $1-3Q^2+1+3Q^2=2$, which is
cancelled by the $bc$ contribution of $-2$. A different possibility to have
\cc\ $+2$ is provided simply by a complex scalar field. We will realise this
as a couple of scalars with opposite signatures, combined into
$(\varphi,\bar\varphi)$ with the operator product
$\d\varphi(z)\d\bar\varphi(w)=-1/(z-w)^2$. This leads to the centreless \emt\
\BE\cT=-\d\varphi\,\d\bar\varphi+\d\cB\!\cdot\!\cC\,.\label{TW}\EE Here $\cB$
and $\cC$ are ghosts of weight $(0,1)$. The other topological generators are
derived from the Landau--Ginzburg considerations \cite{[W-LG]} and read
\BE\cH=-{\ctop+3\over6}\,\cB\cC-{\ctop-3\over6}\,\varphi\,\d\bar\varphi
\label{HW}\EE and
\BE\cQ=\frac{\sqrt{2}}{6}\bigl[(\ctop+3)\d\varphi\!\cdot\!\cB+
(\ctop-3)\varphi\,\d\cB\bigr]\,,\qquad\cG=-\sqrt{2}\,\cC\d\bar\varphi
\label{QGW}\EE

This representation of the topological conformal algebra continues smoothly
to $\ctop=3$, where it, obviously, coincides with what we had in
\req{Tc3}--\req{HQGc3} as a result of taking the classical limit of the
`Kazama--Suzuki mapping' \req{Tsl}--\req{QGsl}. We will return to the
construction \req{TW}--\req{QGW} in section~9.

Curiously, the formulae \req{TW}--\req{QGW} are not the only realisation of
the topological conformal algebra with the \emt\ \req{TW}, in terms of the
specified fields $(\varphi;\bar\varphi;\cB,\cC)$. Another one can be obtained
from the above construction \req{construction1}--\req{G} if we bosonise the
matter through a current $\d u$ and then change the basis of fields to
$(\varphi;\bar\varphi;\cB,\cC)$, introduced via\BE b=e^{Q\varphi}\cC\,,\quad
c=e^{-Q\varphi}\cB\,,\qquad\d
u=-\d\bar\varphi+Q\cB\cC+\frac{1-Q^2}{2}\d\varphi\,,\qquad
I=-\d\bar\varphi+Q\cB\cC-\frac{1+Q^2}{2}\d\varphi\label{secondmapping}\EE
where, as before, $Q=\sqrt{(1-d)/3}$ \ and \
$\d\varphi(z)\d\bar\varphi(w)=-1/(z-w)^2$, \ $\cB(z)\cC(w)=1/(z-w)$. In terms
of the new fields, the \emt\ \req{construction1} takes precisely the form
\req{TW}, but the other topological generators are in no simple way related
to those from eqs.~\req{HW} and \req{QGW}. It is doubtful that the
construction based on \req{secondmapping} would have any meaning beyond an
exercise in bosonisation.

\section{(No) MFF states in Landau--Ginzburg theories}\lvm In this section we
use the construction \req{TW}--\req{QGW} \cite{[W-LG]} for the topological
conformal algebra in order to evaluate the topological/MFF singular states in
the $N\!=\!2$ Landau--Ginzburg theory (see \cite{[M]}--\cite{[Wa]} and
references therein; our analysis pertains, obviously, to {\sl undeformed\/}
Landau--Ginzburg models describing {\sl conformal\/} topogical theories).

The `bosonisation' \req{TW}--\req{QGW} does not refer explicitly to either a
`constituent' matter theory with \cc\ \req{d(c)} or to an $sl(2)$ algebra,
but instead it bears a Landau--Ginzburg interpretation, as explained in
\cite{[W-LG]}. However, the formulae for $\cH$ and $\cQ$ should be handled
with caution, since they involve zero mode of the $\varphi$ field. In
particular, the states that the generators are acting on should be defined
carefully. Since $\varphi$ is OPE-isotropic, an obvious possibility is to
start with states represented as functions of $\varphi$ only (and possibly
the ghosts). Then, a state satisfying eqs.~\req{cpsconditions} (i.e. a chiral
primary state) can be represented by the
operator\BE\Psi_j=\varphi^{6\htop/(\ctop-3)}=\varphi^{2j}\EE (where $j$ is
the $sl(2)$ spin, see eq.~\req{htopk}).

Now it becomes possible to evaluate in terms of the construction
\req{TW}--\req{QGW} the singular vectors constructed above. They turn out to
{\it vanish\/}. The derivation is straightforward, starting from the
definition of the mode action\BE\new\BA{rclcrcl}
(\cL_n\Psi)(w)&=&\oint(z-w)^{n+1}\,\cT(z)\!\cdot\!\Psi(w)\,,&{}&
(\cH_n\Psi)(w)&=&\oint(z-w)^n\,\cH(z)\!\cdot\!\Psi(w)\,,\\
(\cQ_n\Psi)(w)&=&\oint(z-w)^n\,\cQ(z)\!\cdot\!\Psi(w)\,,&{}&
(\cG_n\Psi)(w)&=&\oint(z-w)^{n+1}\,\cG(z)\!\cdot\!\Psi(w)\,,\EA\EE and
evaluating operator products in the integrands. As a sample calculation,
consider the term $(\cQ_{-1}\cG_{-1}\Psi)(w)$ (which is a part of the
level-2 vectors \req{1top2} and \req{2top2}): we find
\BE(\cG_{-1}\Psi)(w)=\oint\cG(z)\Psi(w)=2j\sqrt{2}\,\cC(w)\varphi(w)^{2j-1}
\EE and then,\BE\new\BA{l}\cQ(z)(\cG_{-1}\Psi)(w)={4j\over
k+2}\Bigl[{\varphi(z)\varphi(w)^{2j-1}\over(z-w)^2}+
(k+1){\d\varphi(z)\,\varphi(w)^{2j-1}\over z-w}\\{}+
(k+1)\d\varphi(z)\,\cB(z)\cC(w)\,\varphi(w)^{2j-1}-
\varphi(z)\,\d\cB(z)\,\cC(w)\,\varphi(w)^{2j-1}\Bigr]\EA\EE so that
\BE\new\BA{rcl}(\cQ_{-1}\cG_{-1}\Psi)(w)&=&\oint(z-w)^{-1}\cQ(z)
(\cG_{-1}\Psi)(w) ~=~{2j\over
k+2}\Bigl[(2k+3)\d^2\varphi(w)\,\varphi(w)^{2j-1}\\
{}&{}&{}+2(k+1)\d\varphi(w)\,\cB(w)\cC(w)\,\varphi(w)^{2j-1}-
2\d\cB(w)\,\cC(w)\,\varphi(w)^{2j}\Bigr]\EA\EE By doing similarly the other
terms, we check that they add up to zero for the corresponding values of the
$sl(2)$ spin $j=j(r,s)\equiv(r-1)/2-(k+2)(s-1)/2$.

For the $\ket{{\myrm{MFF}}{r1}}$ vectors, the details are as follows: We
start with singular vectors at levels $r=$ 2, 3, and 4 in the topological
guise, as constructed in section~3. When evaluated on a $\varphi^{2j}$
vacuum, they become, respectively, \BE\ket{\Upsilon^{(2)}_{\{{1\over2},0\}}}=
(2j-1)(k+2)\L(-\d\bar\varphi\,\d\varphi\,\varphi^{2j}+
\d\cB\,\cC\,\varphi^{2j}+
2j(k+3)\d\varphi\,\d\varphi\,\varphi^{2j-2}\R)\,,\label{1}\EE
\BE\new\BA{rcl}\ket{\Upsilon^{(3)}_{\!\{1,0\}}}&=&
2(j\!-\!1)(k+2)\Bigl(-2\d\bar\varphi\,\d\bar\varphi\,\d\varphi\,
\varphi^{2j+1}+ 2j(10+3k)\d\bar\varphi\,\d\varphi\,\d\varphi\,\varphi^{2j-1}-
(k\!+\!2)\d\bar\varphi\,\d^2\varphi\,\varphi^{2j}\\{}&{}&{}+
4\d\cB\,\cC\,\d\bar\varphi\,\varphi^{2j+1}-
2j(10\!+\!3k)\d\cB\,\cC\,\d\varphi\varphi^{2j-1}+
(k\!+\!2)\d\cB\d\cC\,\varphi^{2j}+
(k\!-\!2)\d^2\bar\varphi\,\d\varphi\,\varphi^{2j}\\{}&{}&{} +
2(1-2j)j(k+3)(k+4)\d\varphi\,\d\varphi\,\d\varphi\,\varphi^{2j-3}+
(2-k)\d^2\cB\,\cC\,\varphi^{2j}\Bigr)\,,\EA\EE and
{\tenpt\BE\new\BA{l}\ket{\Upsilon^{(4)}_{\{{3\over2},0\}}}=
\half(2j-3)(k+2)\Bigl(
6(k+1)(k+2)\cB\cC\,\d\bar\varphi\,\d^2\varphi\,\varphi^{2j}-
6(k+1)(k+2)\cB\,\d\cC\,\d\bar\varphi\,\d\varphi\,\varphi^{2j}\\{}-
12\,\d\bar\varphi\,\d\bar\varphi\,\d\bar\varphi\,\d\varphi\,\varphi^{2j+2}+
2(-2+80j-k+22jk)\d\bar\varphi\,\d\bar\varphi\,\d\varphi\,\d\varphi\,
\varphi^{2j}-
14(k+2)\d\bar\varphi\,\d\bar\varphi\,\d^2\varphi\,\varphi^{2j+1}\\{} +
36\d\cB\,\cC\,\d\bar\varphi\,\d\bar\varphi\,\varphi^{2j+2}+
8(1-2j)j(43+23k+3k^2)\d\bar\varphi\,\d\varphi\,\d\varphi\,\d\varphi\,
\varphi^{2j-2}+
2j(k+2)(31+8k)\d\bar\varphi\,\d^2\varphi\,\d\varphi\,\varphi^{2j-1}\\{}-
(k+2)(k+4)\d\bar\varphi\,\d^3\varphi\,\varphi^{2j}+
6(k+1)(k+2)\d\cB\,\cB\,\d\cC\,\cC\,\varphi^{2j}+
4(2-80j+k-22jk)\d\cB\,\cC\,\d\bar\varphi\,\d\varphi\,\varphi^{2j}\\{}+
2(22-7k)\d\cB\,\cC\,\d^2\bar\varphi\,\varphi^{2j+1}+
8j(2j-1)(43+23k+3k^2)\d\cB\,\cC\,\d\varphi\,\d\varphi\,\varphi^{2j-2}\\{}+
2j(k+2)(17+6k)\d\cB\,\cC\,\d^2\varphi\,\varphi^{2j-1}+
28(k+2)\d\cB\,\d\cC\,\d\bar\varphi\,\varphi^{2j+1}-
4j(k+2)(24+7k)\d\cB\,\d\cC\,\d\varphi\,\varphi^{2j-1}\\{}+
(k+2)(k+4)\d\cB\,\d^2\cC\,\varphi^{2j}+
2(-22+7k)\d^2\bar\varphi\,\d\bar\varphi\,\d\varphi\,\varphi^{2j+1}+
8j(24-k-2k^2)\d^2\bar\varphi\,\d\varphi\,\d\varphi\,\varphi^{2j-1}\\{}+
2(k+2)(3k-10)\d^2\bar\varphi\,\d^2\varphi\,\varphi^{2j}+
8(j-1)j(2j-1)(k+3)(k+4)(k+5)\d\varphi\,\d\varphi\,\d\varphi\,\d\varphi\,
\varphi^{2j-4}\\{}+ 2(22-7k)\d^2\cB\,\cC\,\d\bar\varphi\,\varphi^{2j+1}+
8j(-24+k+2k^2)\d^2\cB\,\cC\,\d\varphi\,\varphi^{2j-1}+
2(10-3k)(k+2)\d^2\cB\,\d\cC\,\varphi^{2j}\\{}+
2(1-k)(k-4)\d^3\bar\varphi\,\d\varphi\,\varphi^{2j}+
2(k-4)(k-1)\d^3\cB\,\cC\,\varphi^{2j}\Bigr) \,.\label{3}\EA\EE}

\noindent These formulae abuse the notation, since, strictly speaking, the
singular states $\Upsilon^{(r)}_{\{{r-1\over2},0\}}$ have been defined only
upon the state that now takes the form $\varphi^{r-1}$, while in
\req{1}--\req{3} we have $\varphi^{2j}$ with an arbitrary $j$ \ instead.
Anyway, we observe the vanishing of the above states for the `true' value
$j\!=\!j(r,1)\equiv(r-1)/2$. The same holds true for the Landau--Ginzburg
mapping of the other topological/MFF singular states considered in this
paper. For instance, the second level-2 vector maps, in the same way, into
\BE\ket{\Upsilon^{(2)}_{\{0,{1\over2}\}}}=
(k+2)^2(2+2j+k)\L(\d\bar\varphi\,\d\varphi\,\varphi^{2j}-
\d\cB\,\cC\,\varphi^{2j}\R)\,,\EE which vanishes for the corresponding value
of the $sl(2)$ spin $j=j(1,2)=-\half(k+2)$.

\smallskip

It may be interesting to look more carefully at the states obtained by
dropping the factor $(j-j(r,s))$ in the $\ket{{\myrm{MFF}}{rs}}$ states
evaluated on the $\varphi^{2j}$ vacuum (with $j$ set to $j(r,s)$
afterwards). At level 2, for instance, these read
\BE(k+2)(-\d\bar\varphi\d\varphi\,\varphi+\d\cB\,\cC\,\varphi
+(k+3)\d\varphi\,\d\varphi\,\varphi^{-1})\EE
and\BE(k+2)^2(\d\bar\varphi\d\varphi\,\varphi^{-k-2}-\d\cB\,\cC\,
\varphi^{-k-2})\,.\EE These, as well as the other states (for $rs\leq4$) thus
obtained are BRST-closed, while being at the same time {\it non-polynomial\/}
in $\varphi$.

\section{Concluding remarks}\lvm The fact that the topological conformal
theories share their singular states with the $sl(2)$ theories points to a
close relation between these two classes of theories. This relation
generalises the results of \cite{[GP],[GP1]} on the reduction from WZW to
{\sl minimal\/} models, by `lifting it up' to the $N\!=\!2$ case. The null
states are just {\it equal\/}, while the reduction of the decoupling
equations performed in \cite{[GP]} gets replaced with the decomposition
$\frt\to\frm\oplus\frl\oplus[bc]$ of the topological theory into matter,
Liouville, and ghosts, considered in section~2. It can thus be applied to the
whole algebra, not only to its singular states.  Even from the point of view
of the correspondence between $sl(2)$ and minimal singular vectors,
incorporating the ghosts and the Liouville simplifies it greatly, allowing us
to work at the level of singular states as such, rather than the decoupling
{\it equations\/}\footnote{The algebraic nature of the reduction from
$sl(2)$- to minimal singular states has been revealed in \cite{[GP1]} (I
thank V.~Petkova for pointing this paper out to me). As we see now, the BRST
formalism developed there extends naturally to the $N\!=\!2$ framework.}.

Note that when taking singular states in the MFF form, the associated
decoupling equations do not have the usual form of {\sl differential\/}
equations; this, however, can be restored by using the `Sugawara' form (see,
e.g., eq.~\req{S22}) of the same singular vectors (which must be tantamount,
eventually, to repeatedly using the Knizhnik--Zamolodchikov equations for the
correlators and other related identities elaborated in~\cite{[GP]}).

Exploiting the usual Ward identities for the Sugawara Virasoro generators
$\tLS_m$, we get, for the topological/MFF states, the decoupling equations
implemented by differential operators with coefficients from the Lie algebra
$sl(2)$. Alternatively, we could start with the singular states written in
the topological guise and use the construction for the topological algerba
generators in terms of matter, Liouville, and ghosts.  Here, as has been
noted above, it would make sense to discard all the $bc$ ghost contributions,
since, at the level of decoupling equations, this would correspond to taking
all the fields in a given correlator as chiral primary ones. It is these
`ghost-free' reduced decoupling operators that are related to the Virasoro
constraints introduced in terms of the times \req{Miwatransform}
\cite{Solving,[GS3]}. An interesting question is which form the `hidden'
$sl(2)$ structure (due to the MFF description) takes in these reduced
decoupling operators. The answer can be expected in terms of the formalism
developed recently in refs.~\cite{[Dunkl]}--\cite{[DKV]}. It can be seen in
simple examples (although the argument is by far not complete) that the
currents $J^{\pm,0}$ which enter in these operators on top of the twisted
Sugawara Virasoro generators $\tLS_m$, arrange into the combinations
$$(J_{-n},J_0)_{sl(2)}\leadsto-\sum_{b\neq a}{P_{ab}\over(z_b-z_a)^n}\,,$$
with $P_{bc}=(t_b,t_c)_{sl(2)}\equiv
t^0_bt^0_c-\half(t^+_bt^-_c+t^-_bt^+_c)$, where $(~,~)_{sl(2)}$ is the
Killing form and $t^{\pm,0}$ are the {\it Lie\/} algebra generators, with the
standard notation $t_b^\a$ used for $t^\a$ in the $b$-th position in
$1\otimes\ldots\otimes1\otimes t^\a\otimes1\otimes\ldots\otimes1$ (the RHS is
understood to act on a correlator $\L<\Psi(z_a)\prod_{b\neq
a}\Psi_b(z_b)\R>$). It turns out further that, by imposing a `generalised
symmetry' condition $P_{ab}\sim j_aj_b$, where the $j_b$ are $sl(2)$ spins of
insertions in the correlator, the decoupling operators become exactly those
of the matter + Liouville theory (i.e., the reduced `ghost-free' ones). This
points to the relation with certain other extensions of differential
operators, considered recently in a number of papers
\cite{[Dunkl]}--\cite{[DKV]}. There, our `reduced' decoupling operators have
been encountered in the analysis of the Calogero model and they,
too, appeared as a reduction of more general operators that extend the
differential ones by the group that permutes insertions in the correlator
(see, in particular, ref.~\cite{[DKV]}).  Relation of these operators with
the Knizhnik--Zamolodchikov equations has also been pointed out in
\cite{[FV]}.

\medskip

Since the MFF formula \req{mff} gives the general (modulo the remarks
following \req{jformula}) form of $sl(2)$ singular vectors, by feeding into
it the construction \req{JJJtop} for the currents (or, its `irreducible'
version \req{JJJpara}), we get, formally, a general form of the topological
singular states, in the case when the topological theory is interpreted as
the appropriately dressed matter theory. As to the practical use of these
formulae (which is not at all straightforward), note that in the
$\frm\oplus\frl\oplus[bc]$ interpretation of the topological theory, the
chiral primary states \req{states} can be represented by the following
ghost-independent operators:\BE\Psi_j=e^{j\sqrt{{2\over
k+2}}(v-\phi)}\psi_j\label{state}\EE where $\psi_j$ is a {\it matter\/}
primary state. Now, when evaluating the operator products such as
$J^\pm(z)\cdot\Psi_j(w)$, the fusion of the two exponentials does not produce
a pole, while with respect to the energy-momentum tensors entering $J^-$, the
field $\Psi_j$ has dimension {\sl zero\/}. It is very interesting if other,
more {\sl effective\/}, free-field constructions for the $sl(2)$ current
algebra exist that would make the MFF formulae \req{mff} directly applicable
(i.e. would allow a direct evaluation of,~say,~$(J^-_0)^{r+p(k+2)}$).

The `irreducible' form~\req{JJJpara} of the construction \req{JJJtop} for the
$sl(2)$ currents provides the `invariant' version of the Wakimoto
representation that incorporates a central-charge-$d$ matter theory. Thus any
matter with \cc\ $d\leq1$ or $d\geq25$ can be dressed up into an $sl(2)$ WZW
theory (and, conversely, recovered from the latter via the hamiltonian
reduction).  Another dressing, that of ref.~\cite{[GS2]}, makes up a
topological conformal theory, while singular vectors of the two resulting
theories are identical (see the diagram \req{trianglediagram}).

An interesting question is whether the relation between the topological and
the $sl(2)$ algebras can be understood in the context of the hidden $sl(2)$
symmetry of quantum gravity of ref.~\cite{[Po]}, as, for instance, a
`covariant' (involving ghosts, in particular) counterpart of the `light-cone'
$sl(2)$.

\bigskip

It is a pleasure to thank B.~Gato-Rivera and M.A.~Vasiliev, and, especially,
W.~Lerche for a number of valuable discussions.

\newpage
\def\theequation{A.\arabic{equation}} \setcounter{equation}{0}
\section*{Appendix}\lvm Level 4 provides the most representative of the
examples of singular states considered in the text, since there exists the
`difficult' $\{{1\over2},{1\over2}\}$-state along with the $\{{3\over2},0\}$
and $\{0,{3\over2}\}$ ones. All the three topological singular vectors at
level 4 can be derived from the following `prototype' state {\tenpt\xpt
\BE\new\BA{l}\biggl(36(135-54\ctop+7\ctop^2+30\htop+58\ctop\htop)\cH_{-4}
+3\htop(351-18\ctop-5\ctop^2+246\htop-14\ctop\htop-
96\htop^2)(\cG_{-3}\cQ_{-1}-2\cL_{-3}\cH_{-1})\\
{}+2(1215-486\ctop+63\ctop^2+108\htop+657\ctop\htop-
60\ctop^2\htop-\ctop^3\htop+252\htop^2\\ \hfill{}-18\ctop\htop^2+
14\ctop^2\htop^2+216\htop^3-24\ctop\htop^3-144\htop^4)\cL_{-4}{}\\{}
{}+2(486-459\ctop+156\ctop^2+\ctop^3-585\htop+168\ctop\htop-
47\ctop^2\htop-378\htop^2+18\ctop\htop^2+288\htop^3)\cH_{-3}\cL_{-1}\\ {}
+12\htop(189-54\ctop+\ctop^2+84\htop+28\ctop\htop-24\htop^2)\cH_{-2}\cL_{-2}
\\{}+2(810-81\ctop-12\ctop^2-5\ctop^3+72\htop+414\ctop\htop-
14\ctop^2\htop-144\htop^2-96\ctop\htop^2)\cL_{-3}\cL_{-1}\\ {}-
6\htop(270-30\ctop^2+279\htop+6\ctop\htop+11\ctop^2\htop+6\htop^2+
2\ctop\htop^2-48\htop^3)\cL_{-2}^2\\ {}+
(1458-1593\ctop-108\ctop^2-5\ctop^3-1485\htop+588\ctop\htop+
\ctop^2\htop-306\htop^2-54\ctop\htop^2+288\htop^3)\cQ_{-3}\cG_{-1}\\
{}+3(-405+162\ctop-21\ctop^2+288\htop-
282\ctop\htop+2\ctop^2\htop+168\htop^2+56\ctop\htop^2-48\htop^3
)\cQ_{-2}\cG_{-2}\\ {}+216\htop(3+\ctop+4\htop)(\cG_{-2}\cH_{-1}\cQ_{-1}-
\cL_{-2}\cH_{-1}^2)+9(-135+54\ctop-7\ctop^2+6\htop-46\ctop\htop+48\htop^2)
\cG_{-2}\cL_{-1}\cQ_{-1}\\ {}
+3(-243+18\ctop-7\ctop^2-210\htop-70\ctop\htop-48\htop^2)
(\cH_{-2}\cQ_{-1}\cG_{-1}+2\cH_{-2}\cH_{-1}\cL_{-1})\\ {}+
4(27\ctop+12\ctop^2+\ctop^3-720\htop+276\ctop\htop+
4\ctop^2\htop-252\htop^2-84\ctop\htop^2+144\htop^3)\cH_{-2}\cL_{-1}^2\\ {}+
12(162-18\ctop^2+135\htop+54\ctop\htop+11\ctop^2\htop+
6\htop^2+2\ctop\htop^2-48\htop^3)\cL_{-2}\cH_{-1}\cL_{-1}\\ {}+
2(324-972\ctop+204\ctop^2+28\ctop^3+1287\htop-297\ctop\htop-
275\ctop^2\htop-11\ctop^3\htop\\ \hfill{}+1512\htop^2+72\ctop\htop^2+
64\ctop^2\htop^2+180\htop^3+60\ctop\htop^3-288\htop^4)\cL_{-2}\cL_{-1}^2\\
{}+3(-81+162\ctop-57\ctop^2+288\htop-30\ctop\htop+
22\ctop^2\htop+156\htop^2+4\ctop\htop^2-96\htop^3)
\cL_{-2}\cQ_{-1}\cG_{-1}\\{}-
9(27-126\ctop-5\ctop^2-78\htop+22\ctop\htop+48\htop^2)
\cQ_{-2}\cH_{-1}\cG_{-1}
+108(3+\ctop+4\htop)(2\cH_{-1}^3\cL_{-1}+3\cH_{-1}^2\cQ_{-1}\cG_{-1})\\{}
+(1377-486\ctop-207\ctop^2-4\ctop^3-1458\htop+978\ctop\htop+
56\ctop^2\htop-576\htop^2 {}-240\ctop\htop^2+288\htop^3)
\cQ_{-2}\cL_{-1}\cG_{-1}\\ {}
+6(81-126\ctop-11\ctop^2-6\htop-2\ctop\htop+48\htop^2) (\cH_{-1}^2\cL_{-1}^2+
\cH_{-1}\cL_{-1}\cQ_{-1}\cG_{-1})\\{}+
6(-189+27\ctop+33\ctop^2+\ctop^3-72\ctop\htop-8\ctop^2\htop+
36\htop^2+12\ctop\htop^2)\cH_{-1}\cL_{-1}^3\\ {}
+{1\over6}(-405+3024\ctop-738\ctop^2-120\ctop^3-\ctop^4-3294\htop+
1026\ctop\htop+822\ctop^2\htop\\ \hfill{}+38\ctop^3\htop-4644\htop^2-
72\ctop\htop^2-180\ctop^2\htop^2-648\htop^3-216\ctop\htop^3+
864\htop^4)\cL_{-1}^4\\{}+
3(-189+27\ctop+33\ctop^2+\ctop^3-72\ctop\htop-8\ctop^2\htop+
36\htop^2+12\ctop\htop^2)\cL_{-1}^2\cQ_{-1}\cG_{-1}\biggr)\ket\htop
\label{proto}\EA\EE}

\noindent by simply substituting into it the corresponding relation for the
topological $U(1)$ charge from eq.~\req{cases}. While for the singular
vectors thus obtained we have established their identity with the MFF states,
the state \req{proto}, on the contrary, does not demonstrate any good
behaviour under the mapping \req{QGsl}--\req{Tsl}: the $BC$ modes would not
decouple from the $sl(2)$ currents unless \req{proto} is actually reduced to
one of the singular states \req{vect1c}, \req{vect2c} or \req{vect3c}.

\medskip

In the `classical' case of \tcc\ $\ctop\!=\!3$ the mapping to the $sl(2)$
states degenerates. Nevertheless, we find from \req{proto} the corresponding
singular vectors of the $\ctop\!=\!3$ topological conformal algebra:

{\tenpt\BE\new\BA{l}\Bigl(\cH_{-4}+\half\cL_{-4}+
\frac{5}{6}\cH_{-3}\cL_{-1}+
\half\cL_{-3}\cL_{-1}-\frac{41}{12}\cQ_{-3}\cG_{-1}-\fourth
\cQ_{-2}\cG_{-2}-\fourth\cG_{-2}\cL_{-1}\cQ_{-1}-
\frac{7}{6}\cH_{-2}\cH_{-1}\cL_{-1}\\{}+\frac{2}{3}\cH_{-2}\cL_{-1}^2-
\frac{7}{12}\cH_{-2}\cQ_{-1}\cG_{-1}-\fourth\cL_{-2}\cQ_{-1}\cG_{-1} {}+
\frac{11}{4}\cQ_{-2}\cH_{-1}\cG_{-1}
{}-\frac{19}{12}\cQ_{-2}\cL_{-1}\cG_{-1}+\cH_{-1}^3\cL_{-1}\\{}
-\frac{11}{6}\cH_{-1}^2\cL_{-1}^2+
\frac{3}{2}\cH_{-1}^2\cQ_{-1}\cG_{-1}+\cH_{-1}\cL_{-1}^3-
\frac{11}{6}\cH_{-1}\cL_{-1}\cQ_{-1}\cG_{-1}-\sixth\cL_{-1}^4+
\half\cL_{-1}^2\cQ_{-1}\cG_{-1}\Bigr)\ket0\,,\EA\EE

\BE\new\BA{l}\Bigl(
\cH_{-4}+\half\cL_{-4}+\eighth\cG_{-3}\cQ_{-1}+\fourth\cH_{-2}\cL_{-2}-
\fourth\cL_{-3}\cH_{-1}+\fourth\cL_{-3}\cL_{-1}-
\fourth\cL_{-2}^2-\half\cQ_{-3}\cG_{-1}\\ {}-\eighth\cQ_{-2}\cG_{-2}+
\fourth\cG_{-2}\cH_{-1}\cQ_{-1}-
\eighth\cG_{-2}\cL_{-1}\cQ_{-1}-\half\cH_{-2}\cH_{-1}\cL_{-1}-
\fourth\cH_{-2}\cQ_{-1}\cG_{-1}\\{}-\fourth\cL_{-2}\cH_{-1}^2+
\half\cL_{-2}\cH_{-1}\cL_{-1}+\eighth\cL_{-2}\cQ_{-1}\cG_{-1}+
\frac{3}{8}\cQ_{-2}\cH_{-1}\cG_{-1} {}-\eighth\cQ_{-2}\cL_{-1}\cG_{-1}\\
{}+\fourth\cH_{-1}^3\cL_{-1} {}-\fourth\cH_{-1}^2\cL_{-1}^2+
\frac{3}{8}\cH_{-1}^2\cQ_{-1}\cG_{-1}
-\fourth\cH_{-1}\cL_{-1}\cQ_{-1}\cG_{-1}\Bigr)\ket1\,,\EA\EE}

\noindent and

{\tenpt
\BE\new\BA{l}\Bigl(\cH_{-4}+\third\cH_{-3}\cL_{-1}+\half\cH_{-2}\cL_{-2}+
\half\cG_{-2}\cH_{-1}\cQ_{-1}-\half\cH_{-2}\cH_{-1}\cL_{-1}\\
{}-\fourth\cH_{-2}\cQ_{-1}\cG_{-1}
{}-\half\cL_{-2}\cH_{-1}^2+\sixth\cH_{-1}^3\cL_{-1}+
\fourth\cH_{-1}^2\cQ_{-1}\cG_{-1}\Bigr)\ket3\,,\EA\EE}

\noindent which can then be viewed as a `resolution' of the corresponding MFF
states at $k\!=\!\infty$.

\medskip

Unlike the case with mapping the `prototype' level-4 state \req{proto} to the
$sl(2)$ module, the transformation of \req{proto} to the Landau--Ginzburg
variables of section~9 can be performed and is found to be proportional to
$(2j-3)(1+2j+k)(6+2j+3k)$, thus vanishing for each of the three level-4
values of the $sl(2)$ spin $j$, read off from \req{cases}.

\newpage
\def\NPB{Nucl. Phys. B}
\def\PLB{Phys. Lett. B}
\def\MPLA{Mod. Phys. Lett. A}
\def\CMP{Commun. Math. Phys.}
\def\IJMPA{Int. J. Mod. Phys. A}

\end{document}